\documentclass[longauth]{aa}  

\usepackage{graphicx}
\usepackage{txfonts}
\usepackage{subcaption}
\usepackage{lscape}
\usepackage{placeins}
\usepackage{comment}
\usepackage[switch]{lineno}

\usepackage[colorlinks=true,linkcolor=blue, citecolor=blue]{hyperref}

\usepackage{orcidlink}

\usepackage{pgfplotstable} 
\pgfplotsset{compat=1.18}
\usepackage{csvsimple}
\usepackage[english]{babel}

\usepackage{amssymb}
\usepackage[flushleft]{threeparttable}
\usepackage{makecell}  
\usepackage{soul}

\newcommand{\NH}{$N_{\rm{H_2}}$}
\newcommand{\Gnotn}{$G_{0}/n$}
\newcommand{\mum}{$\mu m$}
\newcommand{\kms}{$\rm km\,s^{-1}$}
\newcommand{\cmpcs}{$\rm cm^{-2}$}
\newcommand{\cmpcc}{$\rm cm^{-3}$}

\newcommand{\CO}{$\rm ^{12}CO$}
\newcommand{\thCO}{$\rm ^{13}CO$}
\newcommand{\CeiO}{$\rm C^{18}O$}
\newcommand{\CseO}{$\rm C^{17}O$}
\newcommand{\HCO}{$\rm HCO$}
\newcommand{\HCOp}{$\rm HCO^{+}$}
\newcommand{\DCOp}{$\rm DCO^{+}$}
\newcommand{\HtwCO}{$\rm H_{2}CO$}
\newcommand{\HNCO}{$\rm HNCO$}
\newcommand{\CthHtw}{$\rm C_{3}H_{2}$}
\newcommand{\CHthOH}{$\rm CH_{3}OH$}
\newcommand{\HthCOp}{$\rm H^{13}CO^{+}$}
\newcommand{\HCN}{$\rm HCN$}
\newcommand{\HthCN}{$\rm H^{13}CN$}
\newcommand{\HNC}{$\rm HNC$}
\newcommand{\HNthC}{$\rm HN^{13}C$}
\newcommand{\NtwHp}{$\rm N_{2}H^{+}$}
\newcommand{\twCN}{$\rm ^{12}CN$}
\newcommand{\DNC}{$\rm DNC$}

\newcommand{\CCH}{$\rm CCH$}
\newcommand{\CCS}{$\rm CCS$}
\newcommand{\CS}{$\rm CS$}
\newcommand{\thSO}{$\rm ^{32}SO$}
\newcommand{\SiO}{$\rm SiO$}
\newcommand{\HCSp}{$\rm HCS^{+}$}

\begin{document}

   \title{Chemical diversity of dense cores in Orion B:\\The role of the environment}

   \author{Helena J. Mazurek\inst{\ref{LUX-Paris}}
    \and Maryvonne Gerin\inst{\ref{LUX-Paris}}
    \and Pierre Gratier\inst{\ref{LAB}}  
    \and Jérôme Pety\inst{\ref{IRAM},\ref{LUX-Paris}} 
    \and Emeric Bron\inst{\ref{LUX-Meudon}} 
    \and Evelyne Roueff\inst{\ref{LUX-Meudon}} 
    \and Antoine Roueff\inst{\ref{UdToulon}}
    \and Ivana Be\v{s}lić\inst{\ref{LUX-Paris}} 
    \and Lucas Einig\inst{\ref{IRAM},\ref{GIPSA-Lab}}
    \and Jan H. Orkisz\inst{\ref{IRAM}}
    \and Pierre Palud\inst{\ref{CRISTAL},\ref{LUX-Meudon}}
    \and Miriam G. Santa-Maria\inst{\ref{CSIC}, \ref{UF}}
    \and Léontine Ségal\inst{\ref{IRAM},\ref{IM2NP}}
    \and Antoine Zakardjian\inst{\ref{IRAP}}
    \and S\'ebastien Bardeau\inst{\ref{IRAM}} 
    \and Pierre Chainais\inst{\ref{CRISTAL}}
    \and Simon Coud\'e\inst{\ref{NAOJ}}
    \and Karine Demyk\inst{\ref{IRAP}} 
    \and Victor de Souza Magalhaes\inst{\ref{NRAO}} 
    \and Javier R. Goicoechea\inst{\ref{CSIC}} 
    \and Annie Hughes\inst{\ref{IRAP}} 
    \and David Languignon\inst{\ref{LUX-Meudon}}  
    \and François Levrier\inst{\ref{LPENS}} 
    \and Franck Le Petit\inst{\ref{LUX-Meudon}}  
    \and Dariusz C. Lis\inst{\ref{JPL}} 
    \and Harvey S. Liszt\inst{\ref{NRAO}} 
    \and Nicolas Peretto\inst{\ref{UC}} 
    \and Albrecht Sievers\inst{\ref{IRAM}} 
    \and Pierre-Antoine Thouvenin\inst{\ref{CRISTAL}}
   }
        
   \institute{LUX, Observatoire de Paris, PSL Research University, CNRS, Sorbonne Universités, 75014 Paris, France \label{LUX-Paris}
   \and LUX, Observatoire de Paris, PSL Research University, CNRS, Sorbonne Universit\'es, 92190 Meudon, France \label{LUX-Meudon} 
   \and Laboratoire d'Astrophysique de Bordeaux, Univ. Bordeaux, CNRS,  B18N, Allee Geoffroy Saint-Hilaire,33615 Pessac, France.
   \and IRAM, 300 rue de la Piscine, 38406 Saint Martin d'H\`eres, France. \label{IRAM} 
   \and Université de Toulon, Aix Marseille Univ, CNRS, IM2NP, Toulon, France. \label{UdToulon}
   \and Instituto de Física Fundamental (CSIC). Calle Serrano 121, 28006, Madrid, Spain. \label{CSIC} 
   \and Department of Earth, Environment, and Physics, Worcester State University, Worcester, MA 01602, USA \label{WORC}
   \and Harvard-Smithsonian Center for Astrophysics, 60 Garden Street, Cambridge, MA, 02138, USA. \label{CfA}
   \and Univ. Grenoble Alpes, Inria, CNRS, Grenoble INP, GIPSA-Lab, Grenoble, 38000, France. \label{GIPSA-Lab} 
   \and Univ. Lille, CNRS, Centrale Lille, UMR 9189 - CRIStAL, 59651 Villeneuve d'Ascq, France. \label{CRISTAL} 
   \and Department of Astronomy, University of Florida, P.O. Box 112055, Gainesville, FL 32611. \label{UF}
   \and Université de Toulon, Aix Marseille Univ, CNRS, IM2NP, Toulon, France,\email{antoine.roueff@univ-tln.fr}. \label{IM2NP} 
   \and Institut de Recherche en Astrophysique et Planétologie (IRAP), Université Paul Sabatier, Toulouse cedex 4, France. \label{IRAP} 
   \and National Radio Astronomy Observatory, 520 Edgemont Road, Charlottesville, VA, 22903, USA. \label{NRAO} 
   \and Laboratoire d'Astrophysique de Bordeaux, Univ. Bordeaux, CNRS,  B18N, Allee Geoffroy Saint-Hilaire,33615 Pessac, France. \label{LAB} 
   \and Laboratoire de Physique de l'Ecole normale supérieure, ENS, Université PSL, CNRS, Sorbonne Université, Université de Paris, Sorbonne Paris Cité, Paris, France. \label{LPENS}
   \and Jet Propulsion Laboratory, California Institute of Technology,  4800 Oak Grove Drive, Pasadena, CA 91109, USA. \label{JPL}
   \and School of Physics and Astronomy, Cardiff University, Queen's buildings, Cardiff CF24 3AA, UK. \label{UC}
   \and National Astronomical Observatory of Japan, National Institute of Natural Sciences, 2-21-1 Osawa, Mitaka, Tokyo 181-8588, Japan. \label{NAOJ}
   }


 
  \abstract
  {
  Prestellar cores are the sites of the earliest stages of star formation. Dust continuum observations are often used to identify and characterize their properties yet only a small fraction of them was observed and studied in terms of their composition and dynamical status. Prestellar cores are often analysed as being analogous to template objects such as L1544 in Taurus, which could create an observational bias if this template is not representative of all possible prestellar cores.
  }
  {
  We explore the chemical diversity of prestellar cores and protostellar cores residing in the Orion B giant molecular cloud selected on their dust continuum emission to provide an unbiased view of their line emission properties and how they vary as function of the core parameters and environment. 
  }
  {
  We make use of the large scale maps of Orion B in 25 molecular lines from which we extract information for a sample of 1001 cores selected using positions extracted from \textit{Herschel} dust continuum observations. The main properties of the core sample are derived using the Principal Component Analysis and additional maps of physical parameters: column density \NH, far-ultraviolet (FUV) radiation field $G_0$ and mean volume gas density $n$. Additional high spectral resolution observations of \CeiO$(1-0)$ serve to evaluate the dynamical status of cores.
  }
  {
  The average line width of the cores is larger than what is typically expected for prestellar cores of closer star forming regions, which suggests that cores in Orion B are subjected to stronger turbulence affecting their stability. The first factor of the PCA analysis explaining the variation of the detected line intensities is the core column density of molecular gas. The second factor explains how the core chemical composition is strictly linked to their environment, which can be traced by the ratio of the external FUV radiation field over the core volume density, \Gnotn. Cold and shielded cores exhibit strong emission of \NtwHp, \CHthOH\ and deuterated species, whereas cores exposed to radiation are devoid of typical core tracers, but exhibit emission of \twCN, \HCOp and HCN. The third factor explaining the core chemical diversity is the mean density along the core line of sight, which is also associated with freeze-out and fractionation signatures.  
  }
  {
  Prestellar cores selected based on their dust emission exhibit a wide range of line emission patterns, which can be related to their intrinsic properties (column density \NH, mean volume density $n$) and environment (presence or lack of FUV). The key parameter that distinguishes cores of different emission patterns is \Gnotn.
  }


   \maketitle
%

\section{Introduction}

Prestellar cores constitute an initial stage of star formation. A small number of individual cores in nearby molecular clouds such as Taurus, Perseus or Ophiuchus have been thoroughly studied through detailed line and continuum observations and subsequent chemical modeling (e.g., L1544 \citealp{2023Jensen}; Barnard~1b \citealp{2016Fuente}; L1689N/IRAS 16293E \citealp{2024Pagani} among other similar cores).
Most of the available information on core properties has been extracted from the analysis of dust continuum maps of molecular clouds, such as the Gould Belt Survey performed by the \textit{Herschel} space Observatory \citep{2010Andre}, and subsequent surveys of more distant clouds performed by ALMA (e.g., ALMA-IMF \citealp{2022Motte}, 
ALMAGAL \citealp{2025Molinari}). These multi-wavelengths imaging programs allowed the extraction of core catalogues with thousands of entries, providing information on the core positions with respect to the cloud structures, as well as statistics on core sizes and masses. In such data sets, cores are defined as local maxima in the column density and flux distribution and extracted using dedicated algorithms (\citealp[see][]{2022Pouteau, 2025Coletta} for the ALMA-IMF data).

However, the gas content of cores and their chemical composition is much less documented. Spectral lines contain key information for studying cores, as line profiles provide insight on the gas dynamics and column densities, while the molecular abundance and excitation are related to the physical conditions \citep{2016Fuente, 2025Grassi, 2025Beslic}. Early searches for gas condensations in Taurus \citep{2002Onishi} used the ground state transition of \HthCOp\ to identify cores within regions of \CeiO\ emission, because this line requires high molecular gas densities and column densities to be detected. Other studies of core compositions were based on a restricted set of lines, chosen for their chemical properties and the proximity of the frequencies to allow a simultaneous acquisition (e.g., \citealp{2021Tatematsu}). Deep spectral line surveys have also been performed at millimetre and submillimetre wavelengths toward a very restricted set of targets chosen for their chemical richness, such as the carbon rich core TMC-1 \citep{2021Cernicharo}, and which may not be representative of the full population of cores. It is clear that the statistical properties of cores spectral line emission remain less well documented than their dust continuum emission.
In this work, we present a statistical study of the molecular line emission of dense cores in the Orion B molecular cloud for a set of lines in the spectral range between 72 and 116~GHz. The source sample and observational data are presented in Section \ref{sec:obs}. Section \ref{sec:vel} presents the main characteristic of the detected spectral lines for the full source sample and discuss the kinematic properties. Section \ref{sec:pca} presents a principal component analysis (PCA) of the molecular data. The results are then discussed in Section \ref{sec:discussion} and summarised in Section. \ref{sec:conclusion}.

\begin{figure*}
    \centering
    \includegraphics[width=1\linewidth]{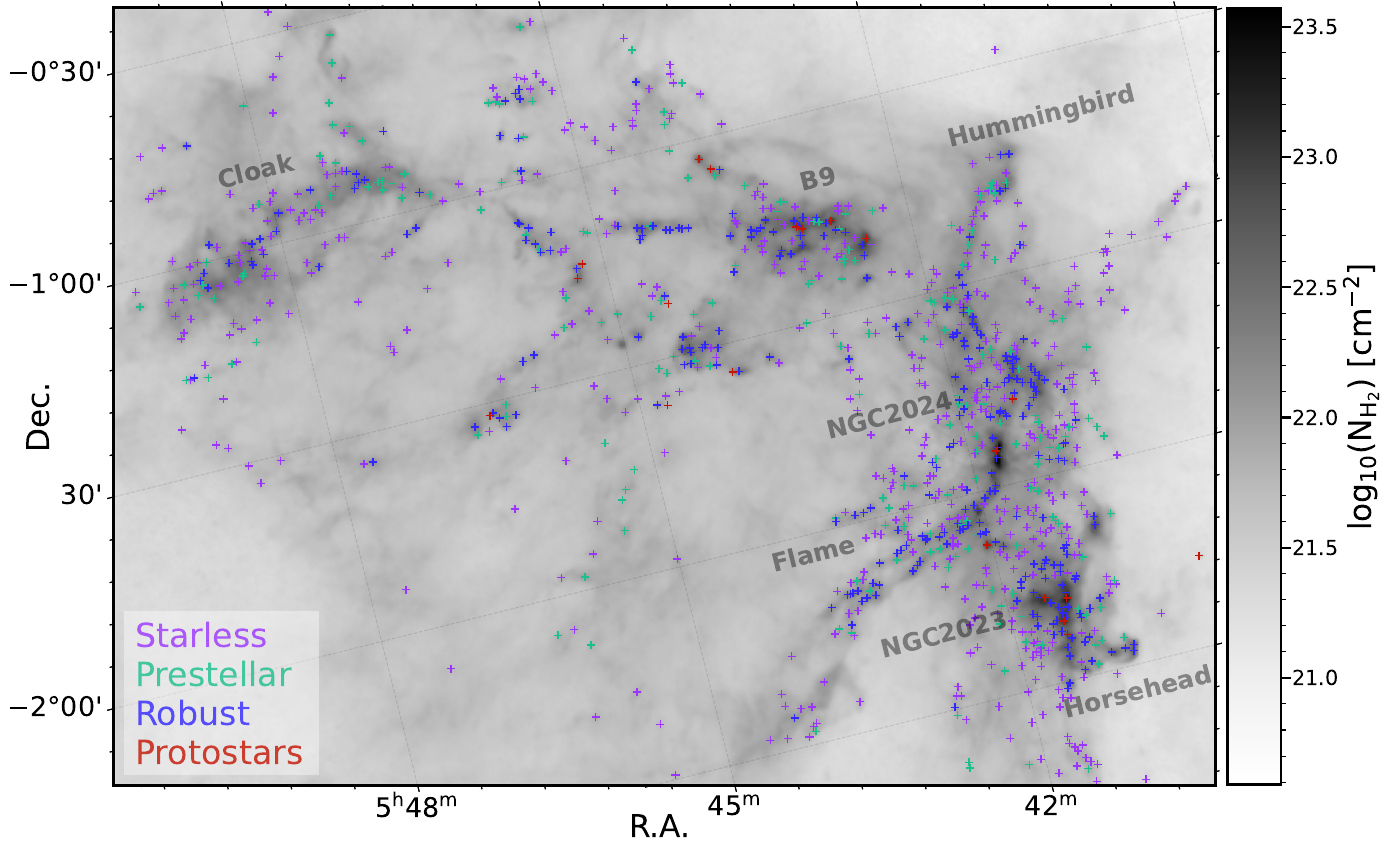}
    \caption{\NH\ column density map across Orion B produced by \cite{2014Lombardi}. Coloured crosses indicate prestellar cores and protostars from the catalogue established by \cite{2020Konyves}}.
    \label{fig:catalog}
\end{figure*}

\section{Observations}
\label{sec:obs}
\subsection{Source catalogue}\label{section:catalog}

The source selection was done based on a catalogue of prestellar cores and protostellar cores by \cite{2020Konyves} (henceforth K20). They have used the \textit{getsources} algorithm \citep{2012Menshchikov} on the multiwavelength \textit{Herschel} observations to decompose the different spatial structures. The progressively larger spatial scales traced by the 160 \mum, 250 \mum, 350 \mum, 500 \mum\ emission along with a high-resolution \NH\ column density map were used to distinguish candidate cores, while additional PACS 70 \mum\ observations served to identify protostellar cores with an internal source producing mid and far infrared thermal emission. 

They identified 1768 starless cores and 76 protostellar cores. Among the starless cores, they identified a subcategory of \textit{robust} cores, defined as those with a mass at least twice the mass of a Bonnor-Ebert core of the equivalent radius and with a kinetic temperature of 10~K. They also extended the sample to include \textit{candidate} prestellar cores selected by adopting a new empirical lower envelope core masses and sizes estimated from their Monte-Carlo core simulations, antecedently used for the survey completeness assessment.

After restricting the catalogue to the Field of View (FOV) covered by the ORION B program, the selection resulted in 1001 objects with 283 robust prestellar, 198 candidate prestellar cores, 500 starless cores and 20 protostellar objects. For each of them, we used the catalogue positions, observed radii, and the core type classification.  
Figure \ref{fig:catalog} shows the positions of the full sample of cores, 
including starless, candidate prestellar and robust cores, and protostars overlayed on a column density map produced by \cite{2014Lombardi}. The different sub-regions within the Orion B cloud are also indicated on the map.

We note here that the positions measured by \cite{2020Konyves} at times do not correspond to the peaks of \NH\ column density maps or to the peaks of the emission of \NtwHp, a commonly used core identifier \citep{1999Lee}. 
However, for the bulk of them we do observe a good correspondence of the positions between the catalogue and data used in our study, and because of this, we decided not to modify the positions even if an offset occurs. This particular concern is discussed further in appendix \ref{app:cores}.   

\subsection{Orion B observations}


\begin{table*}[ht!]
\centering
\caption{Molecular species studied in this work.}
\label{tab:molecular-species}
\begin{tabular}{l|l|l|l|l|l}
\hline\hline
Species & Resolved QNs & Frequency [GHz] & $log(A_{ij})\, [\rm s^{-1}$] & $E_u/k$ [K] & RMS [K] \\ 
\hline
$\rm ^{12}CO$ & $\rm J=1-0$& 115.2712 & -7.14 & 5.53 & 0.13 \\ 
$\rm ^{13}CO$ & $\rm J=1-0$ & 110.2014 & -7.20 & 5.29 & 0.05 \\ 
$\rm C^{18}O$ & $\rm J=1-0$ & 109.7822 & -7.20 & 5.27 & 0.05 \\ 
$\rm C^{17}O$ & $\rm J=1-0$ & 112.3593 & -7.17 & 5.39 & 0.06 \\ 
$\rm HCO$ & $\rm 1(0,1)-0(0,0),J=3/2-1/2,F=2-1$ & 86.6708 & -5.33 & 4.18 & 0.06 \\ 
$\rm HCO^{+}$ & $\rm J=1-0$ & 89.1885 & -4.38 & 4.28 & 0.04 \\ 
$\rm H_{2}CO$ & $1(0,1)-0(0,0)$ & 72.8379 & -5.09 & 3.50 & 0.18 \\ 
$\rm HNCO$ & $4(0,4)-3(0,3),F=5-4$ & 87.9252 & -5.53 & 10.55 & 0.06 \\ 
$\rm C_{3}H_{2}$ & $2(1,2)-1(0,1)$ & 85.3389 & -4.63 & 6.45 & 0.06 \\ 
$\rm CH_{3}OH$ & $2(0,2)-1(0,1)A,vt=0$ & 96.7414 & -5.47 & 6.96 & 0.04 \\ 
$\rm CH_{3}OH$ & $2(1,2)-1(1,1)E,vt=0$ & 96.7394 & -5.59 & 12.54 & 0.04 \\ 
$\rm H^{13}CO^{+}$ & $\rm J=1-0$ & 86.7543 & -4.41 & 4.16 & 0.05 \\ 
$\rm DCO^{+}$ & $\rm J=1-0$ & 72.0393 & -4.16 & 3.46 & 0.20 \\ 
$\rm HCN^*$ & $\rm J=1-0$ & 88.6316 & -4.62 & 4.25 & 0.04 \\ 
$\rm H^{13}CN^*$ & $\rm J=1-0$ & 86.3399 & -4.65 & 4.14 & 0.06 \\ 
$\rm HNC$ & $\rm J=1-0$ & 90.6636 & -4.57 & 4.35 & 0.05 \\ 
$\rm HN^{13}C$ & $\rm J=1-0$ & 87.0909 & -4.73 & 4.18 & 0.04 \\ 
$\rm N_{2}H^{+*}$ & $\rm J=1-0$ & 93.1734 & -4.44 & 4.47 & 0.03 \\ 
$\rm ^{12}CN^*$ & $N=1-0,J=3/2-1/2,F=5/2-3/2$ & 113.4910 & -4.92 & 5.45 & 0.08 \\ 
$\rm DNC$ & $\rm J=1-0$ & 76.3057 & -4.79 & 3.66 & 0.11 \\ 
$\rm ^{12}CS$ & J=2-1 & 97.9810 & -4.78 & 7.05 & 0.04 \\ 
$\rm CCH^*$ & $N=1-0,J=3/2-1/2,F=2-1$ & 87.3169 & -5.82 & 4.19 & 0.06 \\ 
$\rm ^{32}SO$ & $3(2)-2(1)$ & 99.2999 & -4.95 & 9.23 & 0.04 \\ 
$\rm SiO$ & $J=2-1$ & 86.8470 & -4.53 & 6.25 & 0.05 \\ 
$\rm HCS^{+}$ & $J=2-1$ & 85.3479 & -4.95 & 6.14 & 0.06 \\ 
$\rm CCS$ & $N=7-6,J=8-7$ & 93.8701 & -4.43 & 19.89 & 0.07 \\ 
\hline\end{tabular}

\flushleft{\footnotesize{\textbf{Notes.} Species with an asterisk have a resolved hyperfine structure. Here, either only one of the resolved quantum numbers or the non-split number are listed.}}
\end{table*}

In this study, we use the data from the ORION-B project (Outstanding Radio-Imaging of OrioN-B,
co-PIs: J. Pety and M. Gerin, \citealt{2017Pety}). It is an observational program performed with the IRAM 30-meter telescope where a large fraction, 5 degrees squared or 18$\times$13 pc region at a distance of 410 pc (\citealt{2023Cao}), of the southern part of the Orion B giant molecular cloud (GMC) was mapped, with a typical resolution of 26'' or 0.05 pc. The frequency coverage of the observations spans over 40 GHz between 72 and 116 GHz with a spectral resolution of 200 kHz, corresponding to 0.5 \kms. The detailed description of data acquisition and reduction can be found in \cite{2017Pety} and \cite{2023Einig}. The spectral cubes used in this analysis were gridded with 9'' pixels and convolved to the same spectral resolution of 36'', corresponding to the resolution of the cube of the lowest spatial resolution in the available dataset, centred on the emission of \DCOp$(1-0)$. For our analysis, we selected spectral lines that are detected in at least a dozen cores, resulting in a total number of 25 lines. The selected lines include the $1-0$ transition of four CO isotopologues \CO, \thCO, \CeiO \, and  \CseO, of the \HCOp , \HthCOp, \DCOp and \NtwHp molecular ions, as well as low transitions of carbon, nitrogen, and oxygen bearing species, \CCH, \twCN, \HCN, \HNC, \HthCN, \HNthC, \DNC, \HCO, \HtwCO, \CHthOH, \HNCO, c-\CthHtw, of four sulphur bearing species \CS, \thSO, \CCS, \HCSp, and of the shock tracer \SiO \, included in the 72-116 GHz spectral range. A summary of the analysed lines can be found in Tab. \ref{tab:molecular-species}. We additionally use the high resolution ($\delta \nu = 0.1$ \kms; 40 kHz) \CeiO$(1-0)$ data cubes obtained with the VESPA correlator at a spatial resolution of 23.6''.

\subsection{Physical parameters}
Complementary to the observations of the molecular emission, we make use of the \NH\ column density map and $T_{\rm dust}$ dust temperature map from \cite{2014Lombardi}, which were derived from continuum observations of thermal dust emission from the \textit{Herschel} Gould Belt Survey (HGBS, \citealt{2010Andre}; \citealt{2020Konyves}) and \textit{Planck} satellite observations (\citealt{2014Planck}). They performed SED fitting on the \textit{Herschel} data with a modified black body as a base model which produced dust opacity maps at 850 $\mu$m and effective dust temperature maps. The dust opacity maps were then converted to visual extinction using a scaling factor appropriate for Orion B, $A_V (\rm {mag}) = 2.7\times10^4\,\tau_{850}$, and subsequently to $\rm H$ column densities following a conversion $N_{\rm H} = 1.8 \times 10^{21} \times A_V\,\rm [cm^{-2}\,mag^{-1}]$, and finally, assuming that all gas along the line of sight is molecular, to $N_{\rm H_2} = N_{\rm H} / 2$. Due to severe pixel saturation around the NGC~2024 nebula, an additional correction was applied to the affected pixels, by scaling the HGBS data with the \cite{2014Lombardi} data as described in e.g. \cite{2025Orkisz}. 
We also make use of the map of the strength of the incident FUV radiation field expressed as a multiplicative factor of the mean interstellar radiation field $G_0$, derived by \cite{2023SantaMaria} from the far infrared luminosity. Finally, we use the 
mean density map for each pixel derived by \cite{2025Orkisz} and previously used by \cite{2025Beslic} in their derivation of the ionization fraction. We chose the mean mass weighted density to have one representative value of the gas density for each core. 

\subsection{Data analysis}

\begin{figure*}[ht!]
    \centering
    \includegraphics[width=0.85\linewidth, trim=0 3 6 0, clip]{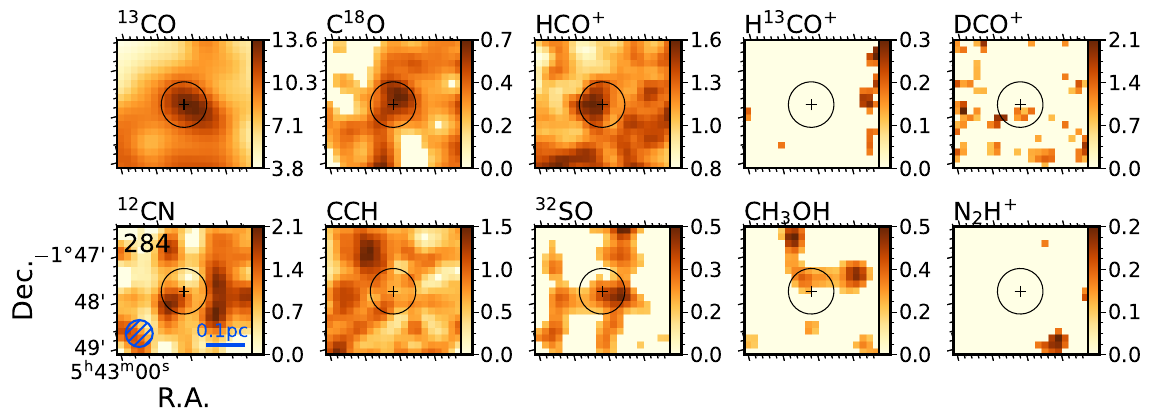}\\
    \includegraphics[width=0.85\linewidth, trim=0 3 6 0, clip]{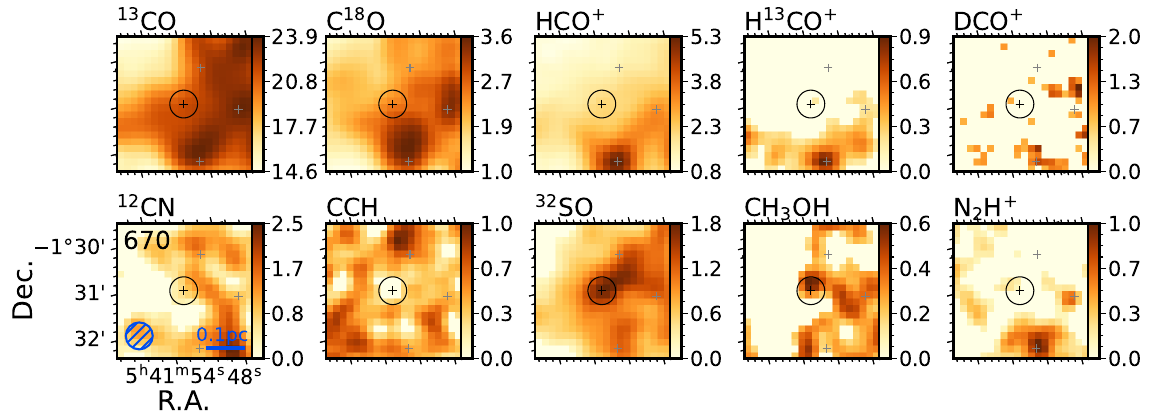}\\
    \includegraphics[width=0.85\linewidth, trim=0 3 6 0, clip]{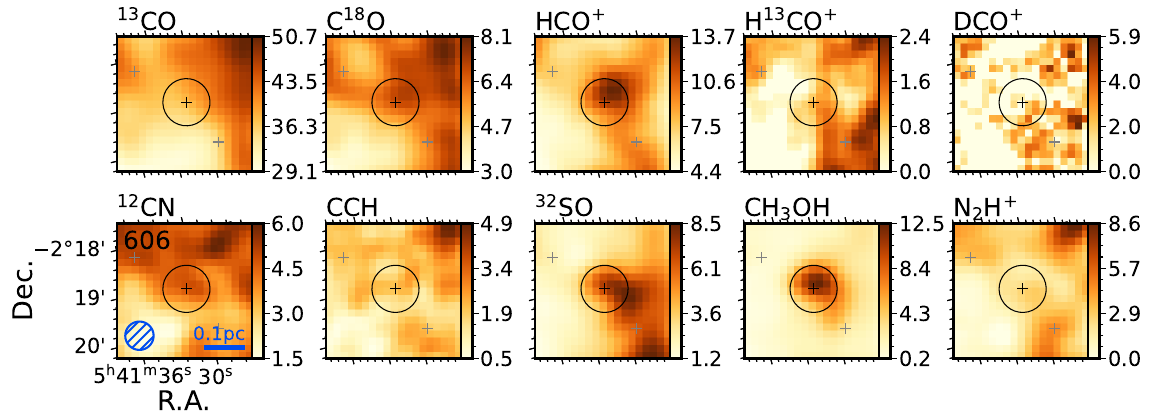}\\
    \includegraphics[width=0.85\linewidth, trim=0 3 6 0, clip]{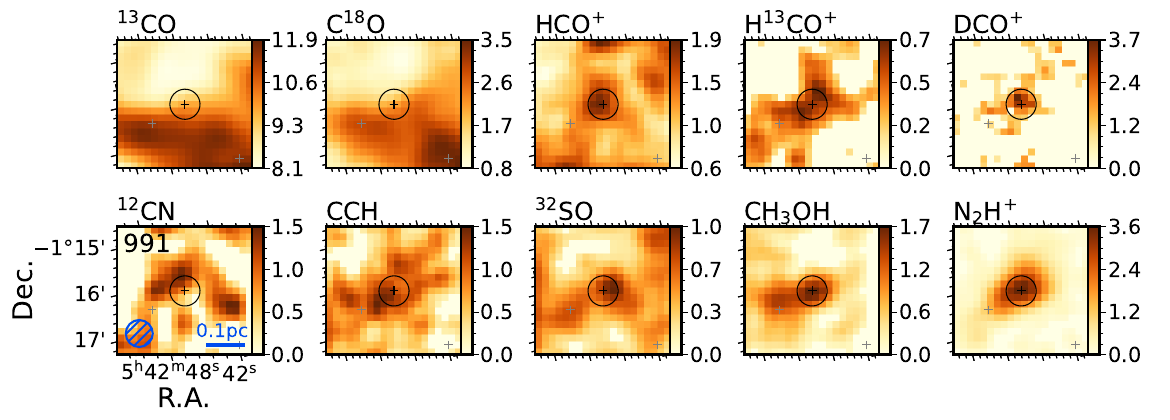}
    \caption{Moment 0 maps for 10 high SNR lines for four examples of objects of the following categories in order: starless (core 284), prestellar (670), robust (606) and protostellar (991), representative of an average column density per category. The species are ordered similarly to Fig. \ref{fig:intensities}, with lines increasingly tracing the extended to dense environments, from left to right. The black circles show the core size as extracted by K20 (see Tab. \ref{tab:cores} for the full list of core parameters). The beam is shown as a blue shaded circle on the lower left panel.}
    \label{fig:moment0-cores}
\end{figure*}

\begin{figure}[hbt!]
\centering
\includegraphics[width=0.5\textwidth, trim = 0 0 0 35, clip]{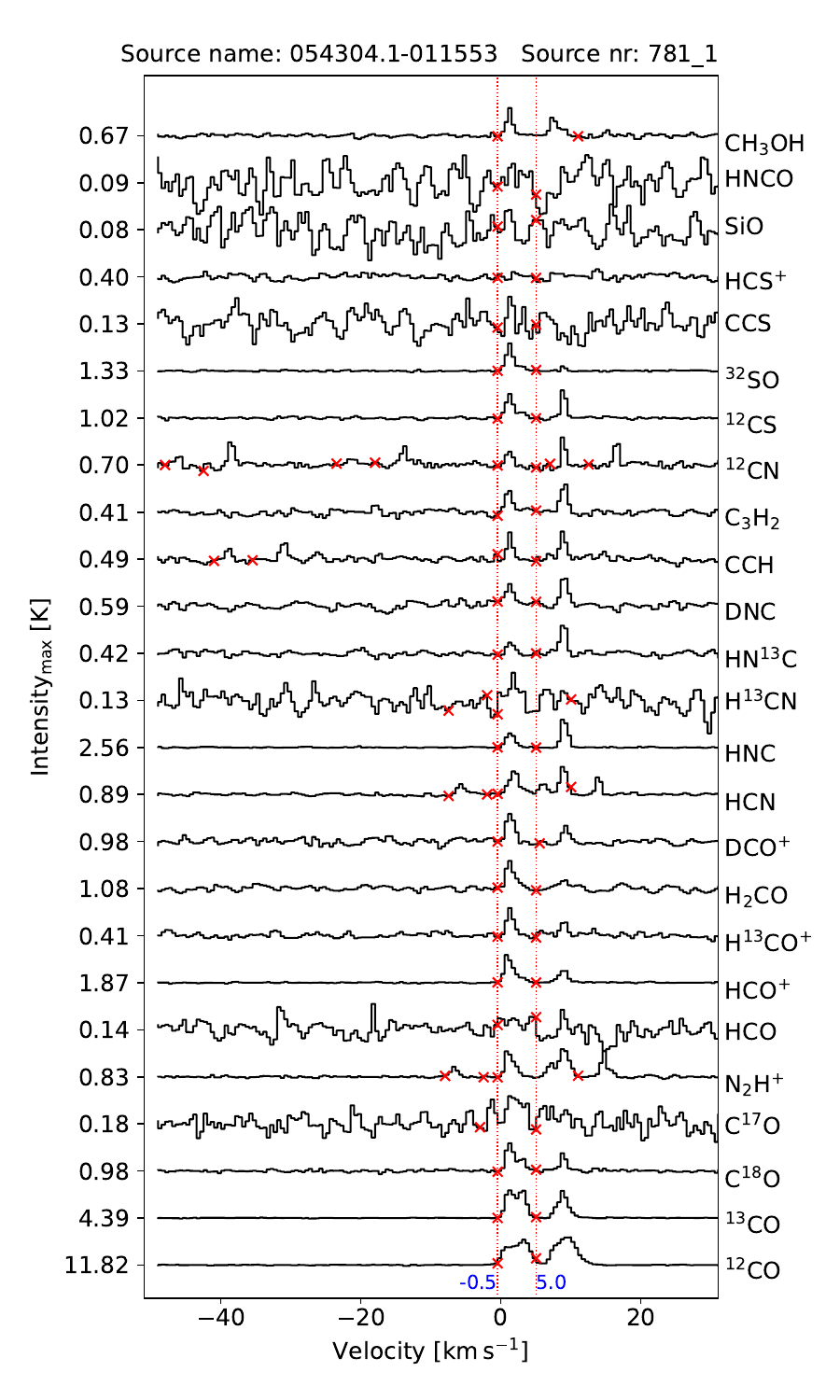}
\caption{Example of the mean spectra extracted for the core 781-1. This particular case also illustrates the ambiguity of a double source as evidenced from the \HthCOp\ profile. The red, vertical dotted line represents the velocity window measured on the SNR cube of \HCOp. Additionally, the red ''x'' symbols show the adjusted windows if multiple, hyperfine components were present in the spectra. All spectra were normalised and the normalisation factor is indicated to the left of the spectra.}
\label{fig:spectra}
\end{figure}

Each object analysed in this work is characterised by a set of fluxes, one for each spectral line. The flux is measured on the spectra averaged over a radius centered on the core position from the K20 catalog. To estimate the velocity windows for each core in a consistent way and to account for both dynamically stable and active environments, a separate cube containing the signal-to-noise (SNR) treated signal of \HCOp$(1-0)$ was used as the input information (cube extraction method described in \citet{2023Einig}. The windows were defined as channels where consecutive signal of \HCOp\ was present in the spectra, and  the same selections of channels were later applied to each molecular line in a consistent manner. The lines with hyperfine structure required a separate definition of spectral windows, to encompass the emission of all hyperfine components using the hyperfine transition frequencies. For cases where window estimation was not possible either due to low SNR or cases where the channel contiguity could not be ensured, a default velocity window was adopted. This default window of (--4 \kms, 16 \kms)  was estimated from the \CeiO\ spectrum averaged over the whole field of view, as \CeiO\ is the closest tracer of the filamentary structures in which the cores reside. Additionally, by verifying the profiles of the \HthCOp\ emission, another reliable core tracer \citep{2003Takakuwa}, which unlike \NtwHp\ has no resolved hyperfine structure, we were able to distinguish between objects laying in different velocity layers. We identified six such cases (marked with an additional digit i.e. Index-1 and Index-2), all residing within the B9 region, and these were considered as separate objects with their own appropriate spectral windows also estimated from the \HCOp\ mask.  This addition extended the sample of core flux measurements to 1007. An example of the extracted spectra with the annotated windows can be seen in Fig. \ref{fig:spectra}. 

Figure \ref{fig:moment0-cores} shows the moment 0 or intensity maps integrated over those same velocity windows for ten high SNR molecular lines toward four selected cores, a starless core 284, a prestellar core 670, a robust core 606, and a protostellar core 991. The relative spatial distribution of molecular emission varies significantly between lines and objects, which is inherently linked to the intrinsic molecule formation processes and molecule excitation. 

The purpose of this study, however, is to investigate in a consistent way how combined information provided by this set of lines can  provide insight on the diversity of cores sampled over the whole cloud structure. That being said, from this point on, each object is characterised only by a set of 25 line intensities and a velocity dispersion measured for \CeiO. We list in Table \ref{tab:cores} the final list of cores studied in this work.

\section{Statistical properties of the core molecular emission}
\label{sec:vel}

\subsection{Line emission overview}
\label{sec:overview}
\begin{figure*}[hbt!]
    \centering
    \includegraphics[width=1\linewidth]{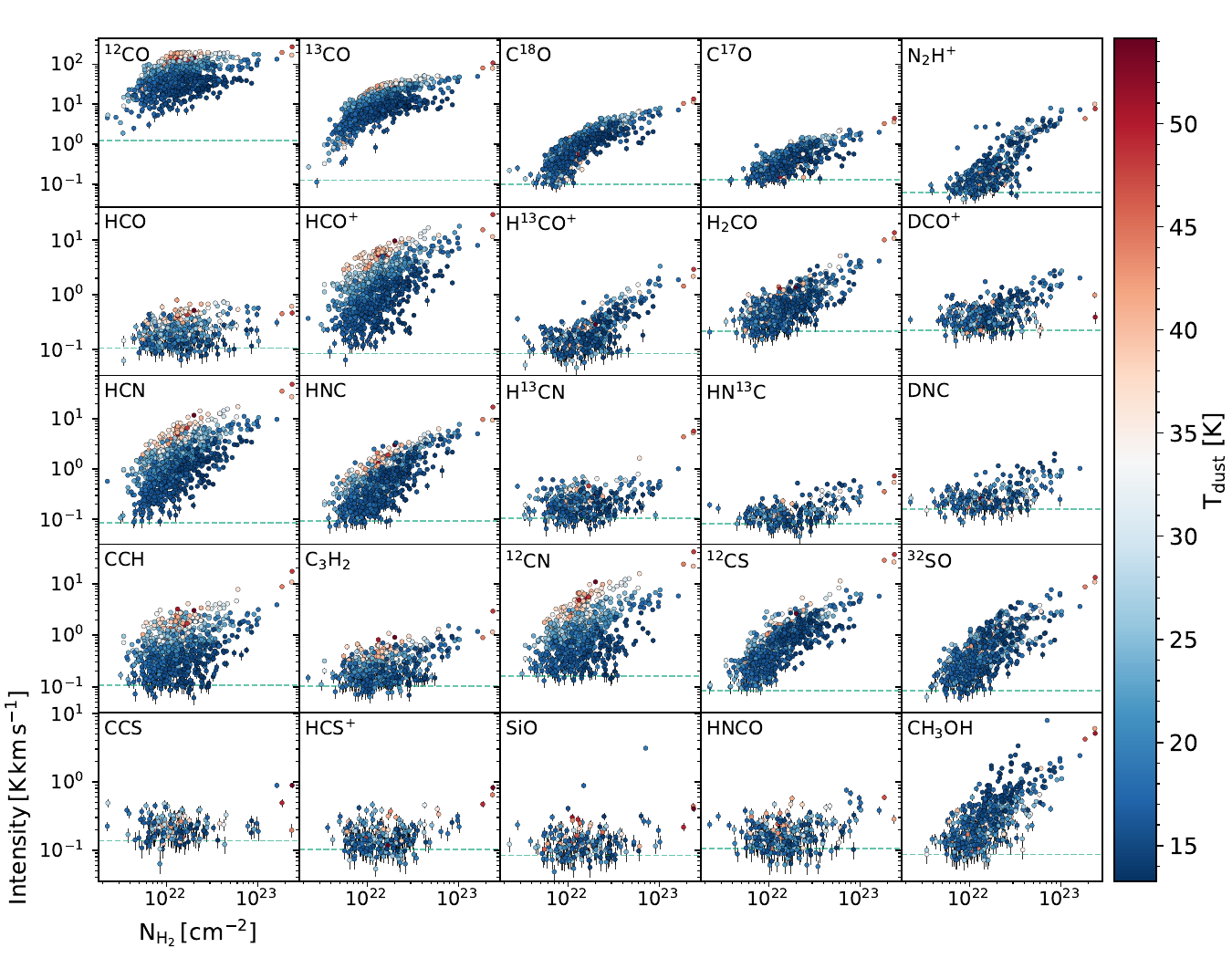}
    \caption{Integrated line intensity as a function of \NH\ column density for all investigated species measured for each source from the 
    sample. The species are either grouped by their "chemical family" counterparts from left to right, or present increasingly denser gas tracers, from UV-illuminated to filamentary and dense core tracers. Particularly, the last column contains typical and often reported in literature dense core tracers. Each point is colour coded with dust temperatures measured per source.
    The horizontal dashed line indicates the median rms measured for each molecular line separately.}
    \label{fig:intensities}
    
\end{figure*}

Figure \ref{fig:intensities} shows the relation between the integrated line intensities and the \NH\ column density as it was measured for each source in the sample. The marker colour corresponds to the dust temperature derived from the SED fitting \citep{2014Lombardi}. The first three rows of the figure show three groups of chemical families: CO, HCO and HCN related families, whereas the species in the last two rows have no direct link with one another, but similarly to the first three rows, are ordered from lines with more extended emission on the left side, tracing translucent and filamentary gas  to localized, dense gas  tracers on the right side. The range of \NH\ values varies from $\rm 2.8\times10^{21}$ \cmpcs\ up to $\rm 2.4\times10^{23}$ \cmpcs \, signifying that the objects trace environments spanning from relatively low $A_V$ of order [3 mag] up to $A_V$ of [240 mag], areas of extremely shielded material. The expected and known correlation between the intensities and \NH\ is visible for almost all examined species. The overall relation between \NH and the line intensity varies significantly from one species to another, a clear sign of their dependency on the environment and underlying conditions of the molecule formation and excitation processes.  There is a visible dust temperature gradient of the integrated intensities for several species, located on the left side of the plot (see notably $^{12}$CO, HCO, HCO$^+$, HCN, CCH and CN). For other species (N$_2$H$^+$, DCO$^+$, HN$^{13}$C, DNC), the emission rises only when \NH\ becomes greater than a threshold of $\sim 10^{22}$ cm$^{-2}$, indicating a sharp change of the abundance and/or excitation conditions when reaching this threshold column density. The relationship between the integrated intensities and \NH\ saturates for the four CO isotopologues while it is almost linear for \NtwHp, \HtwCO\ or \CHthOH. HNC and SO present an intermediate behaviour with a clear increase of the line fluxes with \NH\ but more dispersion around the mean trend. This behaviour must be related to a chemical mechanism as it does not depend on the magnitude of the line intensities themselves which vary across more than one order of magnitude when considering the CO isotopologues. It is also important to note here that the few points which deviate from the main branch (most noticeable for \NtwHp) correspond to cores lying in a part of the B9 region where we register complex velocity structure (cores 301, 786, 993). As we cannot entirely distinguish the different \NH\ layers, a single measurement of the $I$(\NtwHp)/$ N_{\rm H_2}$ is higher. Moreover, the protostar 996 and the core 829, lying at the very centre of the map, also belong to such examples, yet they do not exhibit a complex velocity structure. They might have a unique overabundance of \NtwHp\ as compared to the rest of the cores in the cloud.

Species with marginal detections e.g., \CCS, \HCSp\ or \HNCO, are mainly seen toward sources that are exposed to unique conditions (e.g., shock-affected regions), thus the statistics are too crude to  probe the scaling of these lines with \NH. SiO is a particular case because its abundance is enhanced in shocks \citep{2001Hatchell}. In protostellar cores, SiO emission is associated with molecular outflows and jets. The bulk of the sample does not present any detectable emission , but some objects exhibit relatively bright SiO emission with high velocity wings indicative of molecular outflows. We identified two such cores, one near the NGC~2023 region (core 606) and a protostar (995) to the north-east of B9 region.  

It is interesting to note that the $1-0$ transitions of \CeiO\ and \CseO, the two CO isotopologues with the lowest abundances and the most fragile with respect to FUV radiation, show a very similar pattern with respect to \NH. In fact, \CseO$(1-0)$ intensities follow those of \CeiO$(1-0)$ with a nearly uniform intensity ratio of \CseO/\CeiO\ $ = 0.291\pm 0.001 $ as further discussed in the appendix and illustrated in   Figure \ref{fig:co_iso_ratio}. This close similarity indicates that the two species trace the same environment, and that the \CeiO$(1-0)$ emission does not present a strong saturation effect caused by high opacity, as such an opacity effect would affect the similarity between the two lines. Therefore, both lines can be considered to be optically thin. The strong similarity with little variation of the line intensities ratio also indicates that these lines have similar excitation temperatures, because differences in excitation would translate in different relative intensities. 

Among the CO isotopologues, the C$^{18}$O$(1-0)$ is optically thin with good SNR on most objects. It is detected in most of cores and well correlated with the H$_2$ column density inferred from the dust emission. Its line profile is therefore determined by the gas motions along the line of sight with no need to correct from opacity broadening. For these reasons, we have used high spectral resolution data of the \CeiO$(1-0)$ line to investigate the kinematic properties of the cores.

\subsection{Velocity dispersion and dynamical status}

In order to estimate the line widths for each core, we used the additional \CeiO$(1-0)$ data cubes of higher spectral resolution, allowing to discern the different velocity structures. As before, the average spectrum over the measured core radius was extracted to which a single Gaussian profile was fitted. The \texttt{fit-lines} routine from the \texttt{spectral-cube} package was used for the line profile fits, along with the \texttt{find-peaks} module from \texttt{scipy.signal} library for the fit initialization. The fitting procedure had the following steps: first the presence of spectral lines within the predefined core windows was verified using \texttt{find-peaks}. Then, only the brightest peak and its parameters (position and amplitude) were used to initialize the single Gaussian fit. The uncertainty on the fit parameters (centroid velocity $\mu$, velocity dispersion $\sigma$, and peak intensity $T_p$) was obtained from the covariance matrix calculated from the fit minimization routine. After the fitting process was completed, a small number of bad fits were eliminated while verifying the following conditions for the fit parameters. The cases where the full width at half maximum (FWHM), calculated as $2\sqrt{ln2}\sigma$, exceeded 5 \kms \, were rejected a priori, as they resulted from a fit of baseline undulations instead of the line  profile. 
Cases where the uncertainty on the velocity dispersion was too high, $\delta \sigma>0.2$ \kms, were also rejected, because the fit uncertainty should come close to the intrinsic channel width, and anything larger than $2\,\delta \nu$ can be considered as a poor fit. Finally, cases with an arbitrary uncertainty threshold of $\frac{\delta \sigma}{\sigma} >30\%$ were also rejected, again indicating poor fits. Out of 1007 objects about 831 have a well-fitted profile. A fraction of them, however, shows clear signs of complex velocity structure where two or more narrow lines are superimposed in the line profile. 
It is not necessarily indicative of the presence of multiple objects along the line of sight as only one feature is detected in other lines like \HthCOp$(1-0)$. The fitted widths of such objects are much larger than the average value due to this spectral complexity,  
but they were kept in the sample as such a fit is still reflective of the average velocity structure along the line of sight. 

\begin{figure}
    \includegraphics[width=1\linewidth]{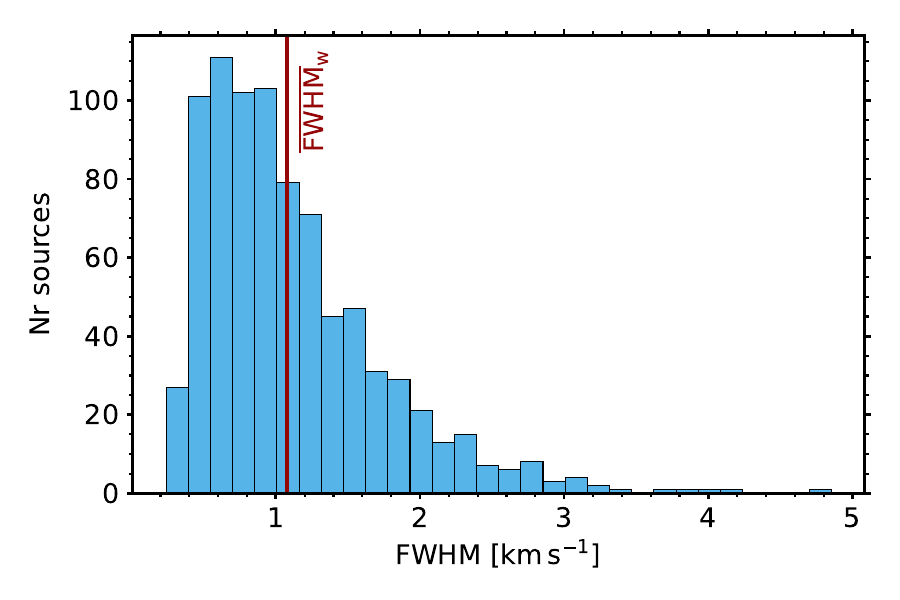}
    \caption{Distribution of the FWHM measured for each core from the Gaussian fitting of the \CeiO$(1-0)$ spectra. The red line shows the weighted mean FWHM of 1.08 \kms.}
    \label{fig:fwhm}
\end{figure}

The distribution of the measured line-widths estimated as the full width at half maximum (FWHM) of the \CeiO$(1-0)$ line is shown in Fig. \ref{fig:fwhm}. The measured FWHM ranges from 0.3 to 3 \kms and the weighted average is 1.08 \kms, with weights estimated as the ratios of the FWHM measurement and its uncertainty. A tail of objects with FWHM > 3 \kms\ corresponds to cases with multiple lines or low SNR. All linewidths are larger than the thermal width for the \CeiO\ molecule in the kinetic temperature conditions encountered in Orion B, $T_{K}$ between 10 and 60~K \citep{2021Roueff}, indicating that the line broadening is due to large scale motions. The mean value is about twice larger than the thermal width of H$_2$ at the gas kinetic temperature of $\sim 20$~K (see \citealt{2006Goicoechea,2024Segal} for cores in the Horsehead nebula), $\sigma_{th} = \sqrt{k_BT/\mu m_H} = 0.29 \sqrt{T/20 K}$ \kms\ for H$_2$, or $FWHM = 0.5  \sqrt{T/20 K}$ \kms. This indicates significant turbulent or systematic motions along the lines of sight to cores. For a single pointed observation of an optically thin line it is not possible to disentangle collapse motions from turbulent motions since both provide similar contributions to the line profile integrated over the line of sight and the core radial extend. Combined information from the variation of the line profile across the core and between an optically thin and an optically thick lines is needed to separate the different types of line broadening \citep{1999Lee}. In general, the velocity structure of the Orion-B cloud is complex with three main velocity layers and velocity structures within each layer \citep{2023Gaudel}. For cores at the low end of the FWHM distribution, the line profiles are nearly consistent with the thermal velocity dispersion for H$_2$ if the kinetic temperatures are between 10 and 20~K. As Orion B is a massive star forming region experiencing intense FUV and dynamical feedback, such a higher kinetic temperature and velocity dispersion as compared with the quiescent environment of Taurus with a kinetic temperature of about $10$~K \citep{2002Onishi} is expected.

\begin{figure}
    \includegraphics[width=1\linewidth]{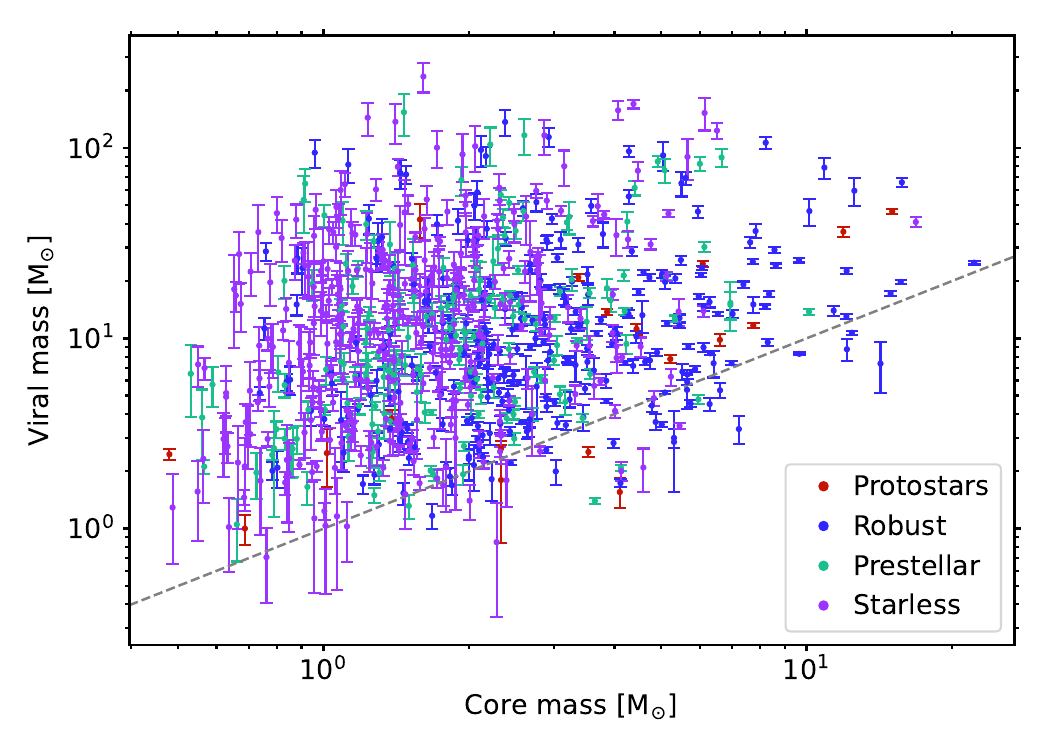}
    \caption{Virial masses as a function of core gas masses measured for all  cores with a high spectral resolution \CeiO$(1-0)$ line profile. The robust cores are plotted in blue, the prestellar cores in green, the starless cores in magenta and the cores hosting protostars in red. The dashed gray line shows the 1:1 ratio.}
    \label{fig:virial}
\end{figure}

We used the information on the velocity dispersion to compute the core virial masses as $M_{Vir} = 5\frac{\sigma^2 R}{G}$ \citep{1992Bertoldi} where $R$ is the core radius and $G$ the gravitational constant. The virial masses are shown in 
Fig. \ref{fig:virial} as a function of the core gas masses deduced from the dust derived \NH\ column densities. The colour of the symbols indicates the core category: starless cores in magenta pink, prestellar cores in cyan green, robust cores in blue and protostellar cores in red. For most of the cores, the virial mass is larger than the gas mass deduced from the dust continuum emission as already noticed by \citet{1992Bertoldi}. The virial masses are also larger than the critical Bonnor-Ebert masses calculated by K20 assuming no turbulence and a kinetic temperature of 10~K as $M_{BE,crit} = 2.4\,R_{BE}\frac{c_s^2}{G}$ where $c_s$ is the sound speed, $G$ the gravitational constant and $R_{BE}$ the radius. The robust cores, defined as those having a gas mass at least twice larger than their critical Bonnor-Ebert mass, have higher gas masses than the both the starless cores and the prestellar cores. The robust cores seem to be closer to the virial equilibrium than the other two, less massive categories but overall there are no strong differences in the velocity dispersion and virial masses between the starless, prestellar and robust cores. The criteria used by K20 for identifying the robust cores, most likely to form stars, are based on well studied cold cores in  nearby molecular clouds like Taurus and Perseus. Cores in a massive molecular cloud hosting massive stars are more diverse; their kinetic temperature is higher than 10~K  at the scales probed by our data, and their velocity dispersion is transonic or supersonic.

\section{PCA analysis}
\label{sec:pca}

Having many molecular tracers of different families and varying detectability, the goal of this work is to interpret the potential correlations between these lines, with the ultimate goal of finding the main underlying physical and chemical parameters and processes that  would describe the trends seen for cores in a general and statistical way. For this reason, we decided to apply the Principal Component Analysis (PCA) to the fluxes measured on cores. 

Principal component analysis is a dimensionality reduction technique which transforms the data $X$ into a new, orthogonal coordinate system with the axis defined by the principal components (PC). These components are defined as the successive axes of largest variance in the data, and so for a multidimensional dataset, these are equivalent to the eigenvectors with the highest eigenvalues of the data covariance matrix. The subsequent components are orthogonal to the former, while their projected variance decreases. Based on the principle of finding these PCs, dominant axis of variation in the data are identifiable and thus can be used to reduce the dimension of the input by taking only the most predominant components, or they can serve to understand the driving factors of data variability/dispersion. 

For a given dataset $X$ of shape $(N, k)$, where $N$ represents the number of measurements or samples with $k$ features, the PC space is defined by $p$ eigenvectors associated with the $p$ largest eigenvalues. The correlation between the original data and the PCs is what is called a PC loading. It describes the contribution of the features to the components and is represented by a matrix of shape $(p, k)$. The PC scores are the projection of the original data $X$ onto the PC space and their matrix has a shape of $(N, p)$. 

In the most commonly used PCA (first implemented by Hotelling, 1933) the PCs are calculated by Singular Value Decomposition (SVD) of the centred data $\mathbf{X_c}$. However, when dealing with data that have strongly varying values and scales of the features, it is also necessary to standardize the data i.e., normalize the centred data to unit variance, which ultimately brings the calculation to an estimation of the eigenvectors of the data correlation matrix \citep{2016Jolliffe}. It is this implementation of the PCA that we use in this work, which is available directly in the Python library \texttt{sklearn} \citep{2011-scikit-learn}.

\subsection{Data selection and preparation}

\begin{figure*}[hbt!]
    \centering
    \includegraphics[width=0.95\linewidth]{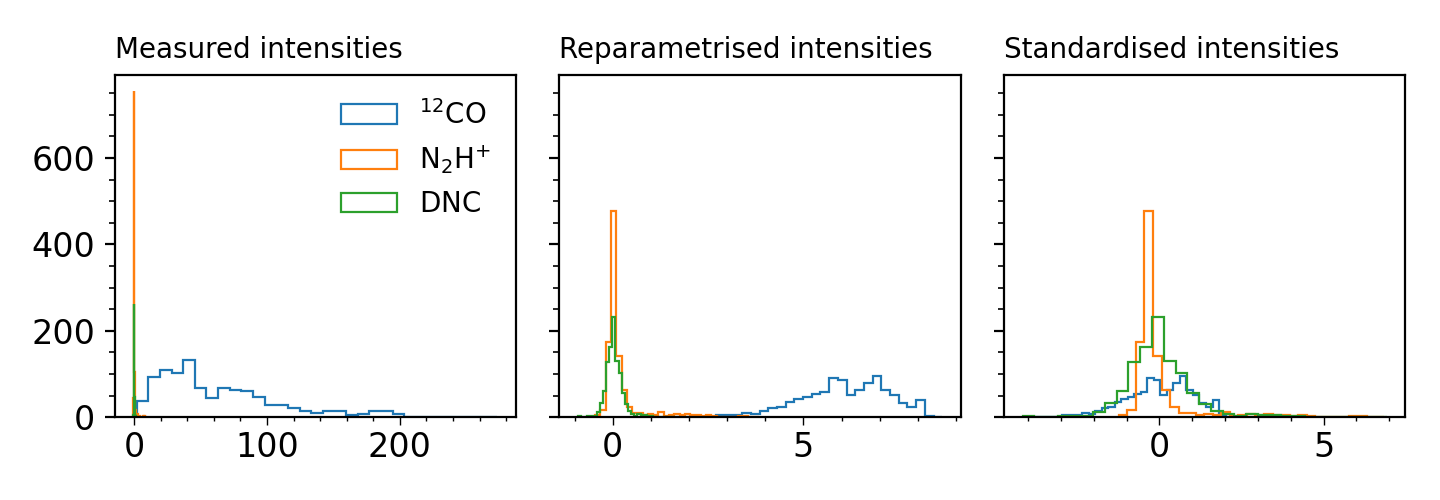}
    \caption{Distribution of the original intensities, reparametrised and standardised intensities for three, selected lines: \CO$(1-0)$, \NtwHp$(1-0)$ and \DNC$(1-0)$.}
    \label{fig:distribution-intensities}
\end{figure*}

Following the methodology described in \cite{2017Gratier}, the data had to be preprocessed to account for the wide dynamical range of intensities and their varying intensity distributions. Example of such distributions for three selected lines are shown in Fig. \ref{fig:distribution-intensities}. They are reflective of the detectability and line strengths which are affected by the conditions at which these molecules form and are collisionally excited, and therefore ought to be different. The inverse hyperbolic sinus function $\rm asinh(x)$ explored by \cite{2017Gratier} to rescale the data was used similarly in this work. It ensures that the low values remain almost unchanged and the high values approach their logarithm leading to a more compact distribution of the reparametrised intensities distribution as illustrated in Fig. \ref{fig:distribution-intensities}.

After the reparametrisation, the intensities were standardised by removing the mean values and dividing them by the standard deviation, both measured for each line of the sample. As a final preparatory step, cores with missing information i.e. for which observations of \DCOp, \HtwCO, \DNC \, and \CCS, had missing values (cores 1, 2, 3 and 982) were removed from the sample. The data prepared as such were passed on to the PCA procedure, with $N=$ 1003 samples of preprocessed intensities and $k=$ 25 features corresponding to the 25 studied molecular lines.

\subsection{Outlier sources}

\begin{figure*}[hbt!]
    \centering
    \includegraphics[width=\linewidth]{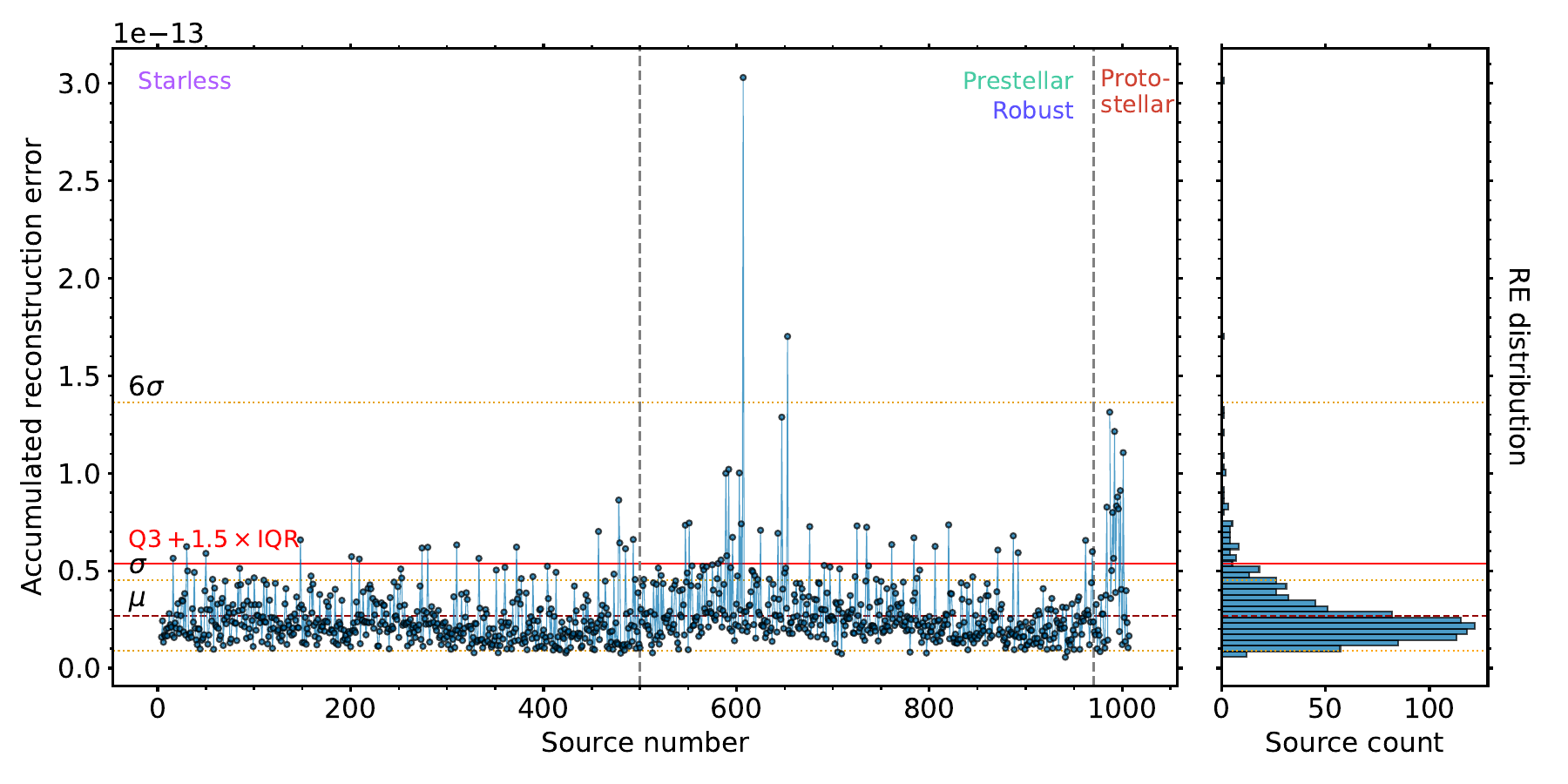}
    \caption{Reconstruction error combined across all features as a function of the source number excluded for a given LOO-PCA trial.
    The right panel shows the distribution of aRE values. The horizontal, dashed lines indicate the mean value $\mu$ (in red), the standard deviation $\sigma$, and a threshold of $6\sigma$ (in yellow) indicative of the strongest outliers. The red horizontal line indicates the upper outlier bound. The gray vertical dashed lines indicate the separation between the sub-classes of cores, which follow the source number/index.}
    \label{fig:pca-RE}
\end{figure*}

In order to probe how robust the analysis is with respect to the sample and, at the same time, to identify the outlier sources, we performed a leave-one-out (LOO) analysis by working on $N$ subsamples in which only one object ($\mathbf{x_{out}}$) was removed. The process consisted of performing the PCA on the $N-1$ sources ($\mathbf{X_{\textit{N-1}}}$) and projecting the excluded source onto the PC space created by the trial sample and transforming it back to the original data space ($\mathbf{x_{out}^{\prime}}$). While the eigenvectors are determined only up to a sign, they can vary across LOO trials. We ensured that they are consistent before each trial by aligning the signs of the PC from the trial sample with the signs determined from the PCA over the full sample. Additionally, both the trial sample and the removed sample were standardised with the mean and variance calculated on the trial sample. This step assured that the properties of the trial sample (mean flux and variance per feature) did not contain any information on the left-out sample. A reconstruction error was calculated for each trial as a squared error $RE =\mathbf{(x_{out}^k} - \mathbf{x_{out}^{\prime\, k})}^2$ for each $k$ feature. Then, an accumulated reconstruction error (aRE) was calculated as the Euclidean norm. It is shown in \ref{fig:pca-RE} as a function of the number of the object removed from the sample. Along with aRE, an uncertainty of the PC loadings due to varying sample was estimated as the standard deviation of the loadings measured in each trial. 
Despite the reparametrisation, the sample contains unique objects, or outliers, which are identifiable through the LOO process. The distribution of aRE is a skewed distribution where its tail represents objects of the strongest peripheral properties or having erroneous measurements. Values deviating from the normal distribution can be assessed by calculating the interquartile range and the upper bound, an outlier detection method introduced by \citet{1977Tukey}. We can report that 53 objects lie above the upper bound, with the two strongest ($> 6\sigma$) identified with core numbers 606, 652. The first of the two, core 606, is a source with a strong and extended outflow observed in \SiO\ and \CHthOH; there is only one such object in the entire Orion B field of view. Then, core 652 lies in the very centre of NGC~2024, representing a region of highest column and volume densities with massive star formation. These objects have the strongest effect on the variability of the principal components. A fraction of outliers correspond to negative fluxes, likely indicative of an erroneous measurements due to worse data quality. We find that the first three PC components are the most robust with respect to a change of the sample, while starting from PC4, the loadings become sensitive to the presence (or absence) of the outliers, as shown in Figure \ref{fig:loading_app0}. 

It is known that PCA is sensitive to outliers as it is based on variance and points which diverge from the principal axis of variation in the data shift it away from the true direction of variance. These outliers certainly disrupt the analysis when all the PCs in the analysis are considered, however, as we find that the first three PCs are not sensitive to them, they were kept in the analysis.

\subsection{PCA results}

\begin{figure}
    \centering
    \includegraphics[width=1\linewidth, trim=0 0 0 0, clip]{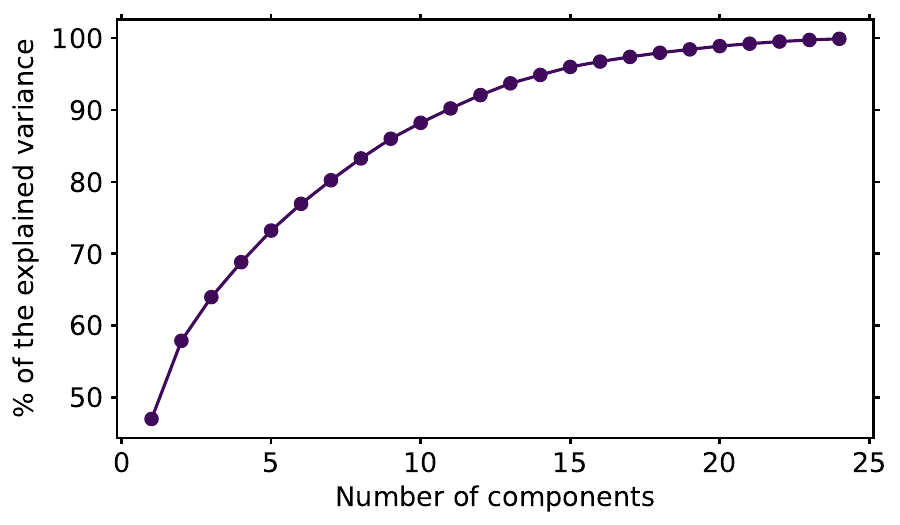}
    \caption{Percentage of the cumulative explained variance ratio as a function of the number of the components considered per PCA execution. Figure modified to adjust font sizes.}
    \label{fig:pca-variance}
\end{figure}

\begin{figure*}
    \centering
    \includegraphics[width=0.33\linewidth]{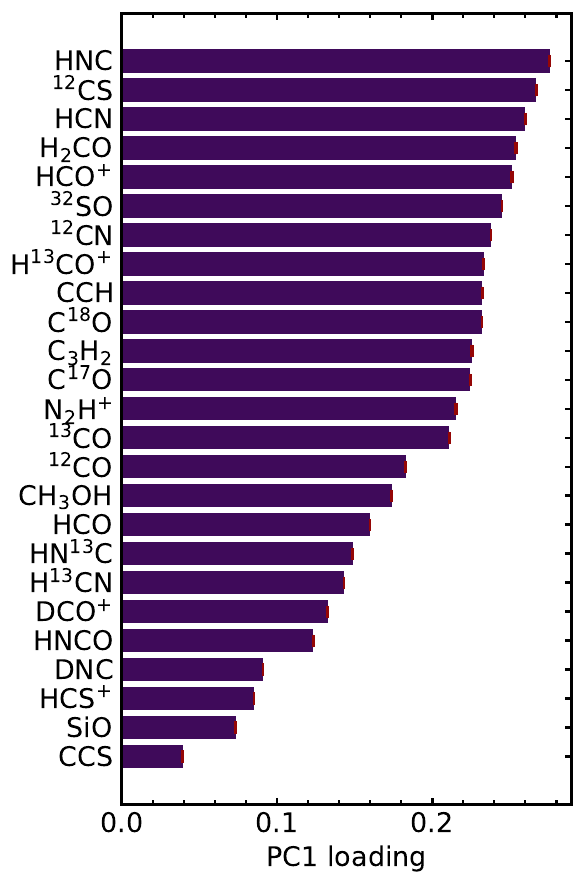}
    \includegraphics[width=0.33\linewidth]{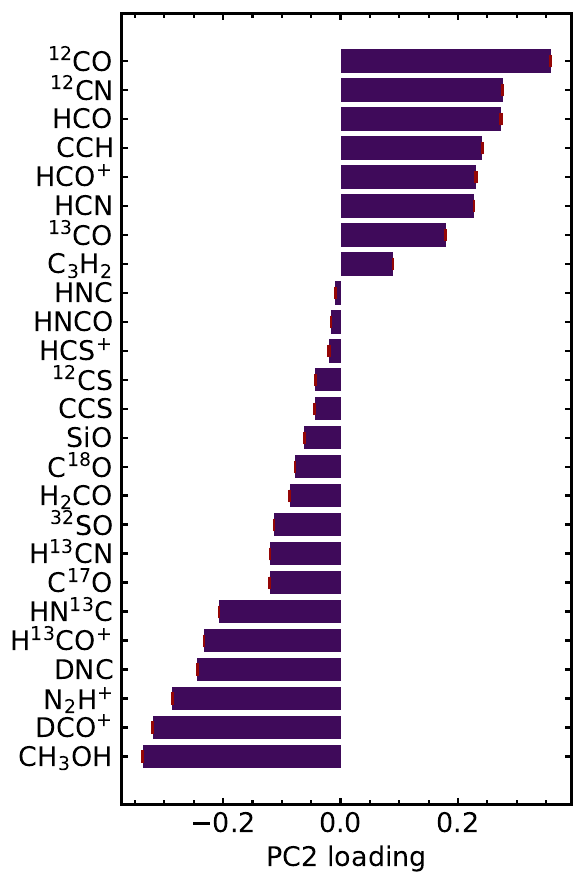}
    \includegraphics[width=0.33\linewidth]{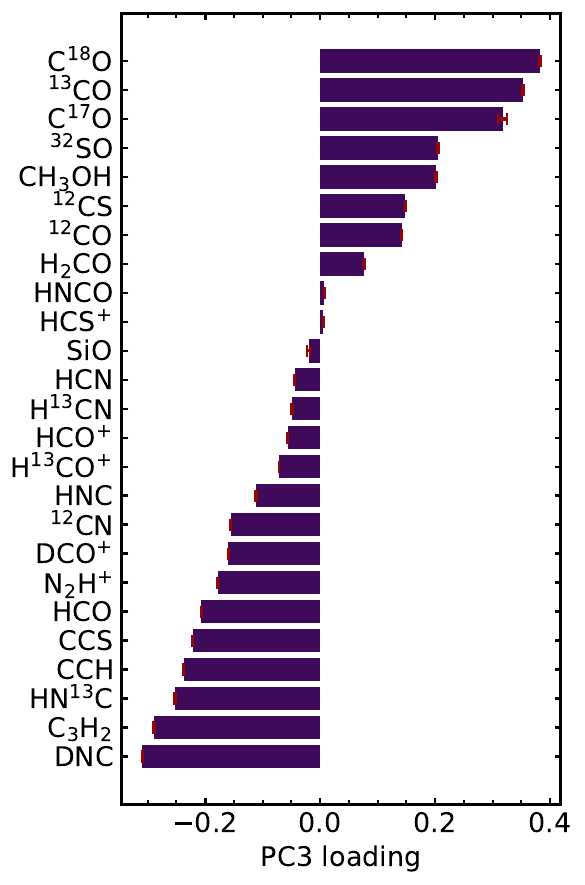}
    \caption{Feature contribution for the first three components, ordered by amplitude of the contribution. Uncertainties measured in the LOO PCA are shown in red.}
    \label{fig:pca-loadings}
\end{figure*}

Figure \ref{fig:pca-variance} shows the cumulative explained variance ratio (EVR) conditional on the number of components, which indicates how much additional information can be obtained by adding further components until all possible are exhausted, which in this case is 25 components. Figure \ref{fig:pca-loadings} presents the feature contribution per component, ordered by amplitude of the contribution.

\begin{figure*}
    \centering
    \includegraphics[width=0.33\linewidth]{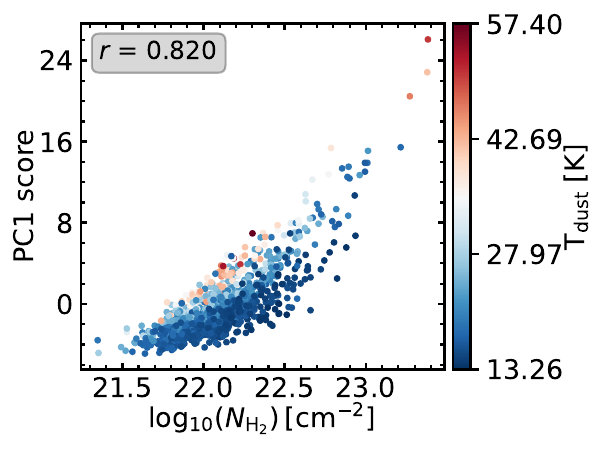}
    \includegraphics[width=0.33\linewidth]{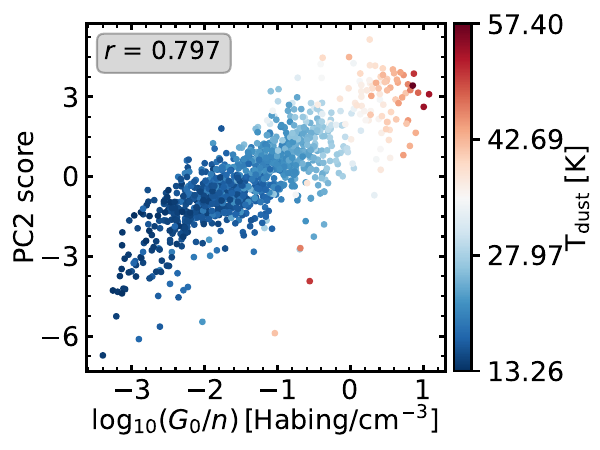}
    \includegraphics[width=0.33\linewidth]{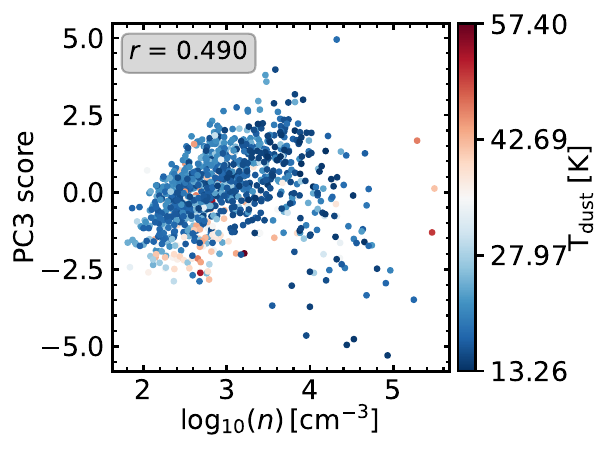}
    \caption{Variation of the PC score as a function of a core physical parameter. The left panel displays PC1 as a function of the H$_2$ column density, the middle panel displays PC2 as a function of the ratio of the FUV field intensity divided by the mean gas density \Gnotn, and the right panel displays PC3 as a function of the mean gas density $n$. Each core is colour coded according to its dust temperature from 13.26~K in dark blue to 57.4~K in dark red. Pearson's correlation coefficient between the parameters is shown in the upper left corner. The correlation coefficient between $n$ and PC3 scores was calculated on densities $<10^{3.5}$ \cmpcc.}
    \label{fig:pca-scores}
\end{figure*}

\begin{figure}[ht!]
    \centering
    \includegraphics[width=1\linewidth]{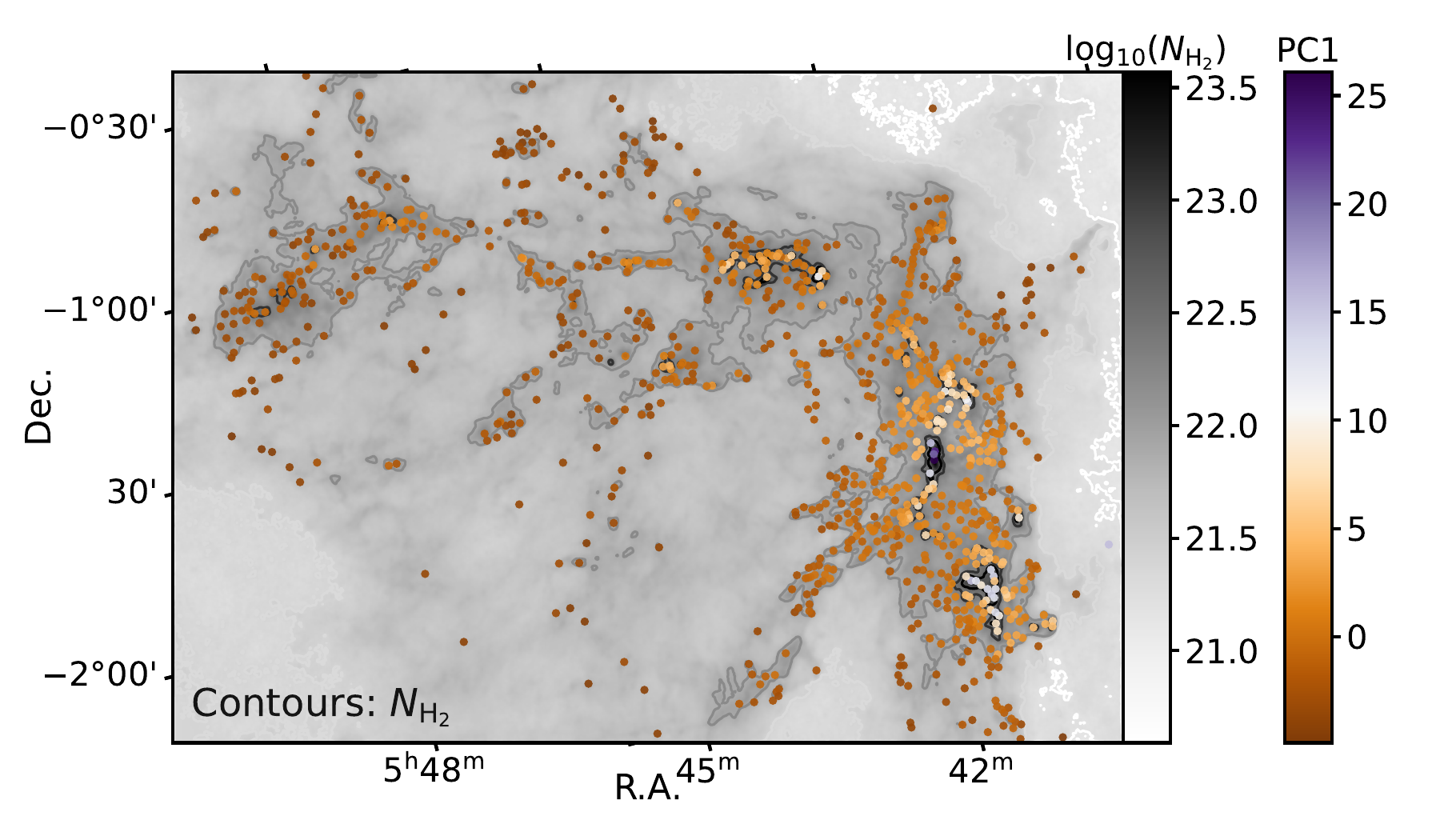}\\
    \includegraphics[width=1\linewidth]{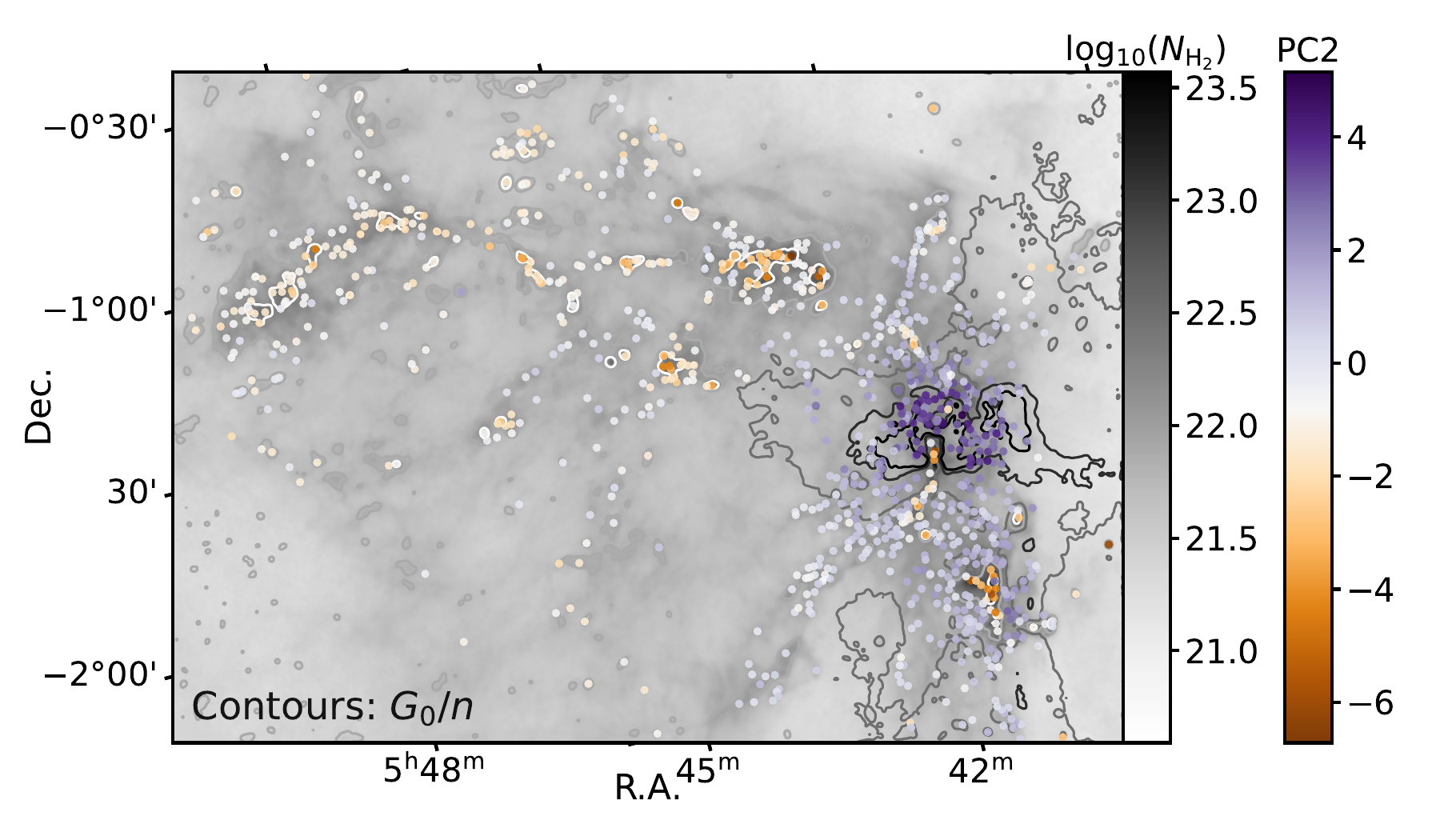}\\
    \includegraphics[width=1\linewidth]{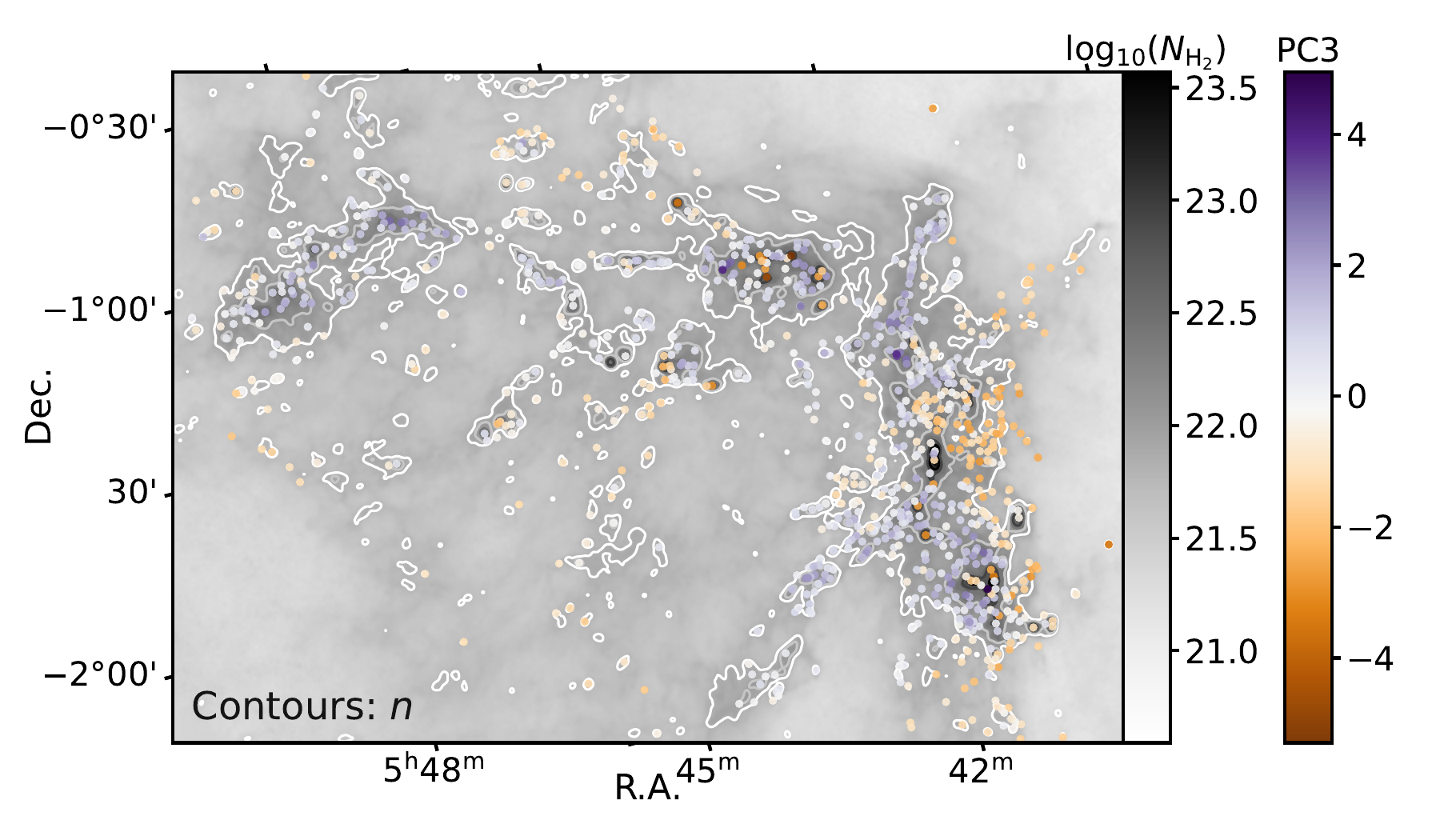}
    \caption{PC scores of cores overlaid on the \NH\ column density map for the first three PC components. Contours correspond to the three physical parameters: \NH, \Gnotn\ and $n$ (smoothed to a 36'' resolution), with the following contour levels [1.5e+21, 2.5e21, 7.5e+21. 2.5e+22, 5e+22] for \NH, [5.0e-03, 5.0e-02, 2.5e-01, 1.5e+00, 5.0e+00] for \Gnotn, and [2.0e+02, 7.0e+02, 4.0e+03, 1.2e+04, 5.0e+04] for $n$.}
    \label{fig:pca-maps}
\end{figure}

We focus on the first three components, PC1, PC2 and PC3, that together represent $\sim$ 65\% of the sample variance. Figure \ref{fig:pca-scores} shows the variation of the PC scores as a function of the logarithms of the selected physical variables: the column density of molecular gas \NH, the ratio of the FUV radiation field intensity divided by the gas density \Gnotn, and the gas density $n$. These three physical parameters were selected after initial assessments of the input fluxes and parameter maps. As described in Section \ref{sec:overview}, the relation between the line fluxes and \NH\, appeared to be the primary observable axis of variation. Based on the combination of lines and the distribution of PC2 scores in Fig. \ref{fig:distribution-intensities} we saw that the environmental impact must have an important role on the second axis of variation. Then, PC3 scores either pointed to dense structures (filaments) or FUV radiation affected regions hinting at direct density effects. Figure \ref{fig:pca-maps} presents the spatial distribution of the PC scores across the Orion B FOV. For each of the components, the core positions are displayed on the \NH\ map, with contours corresponding to the physical parameters \NH, \Gnotn\ and n. The colour of the points corresponds to the PC scores.
    
As noticed in \cite{2017Gratier}, where similar analysis was performed over all cloud scales but a smaller region, the first component only has positive loadings. Here, it explains 47\% of the total variance and represents the general trend of increasing emission with increasing material along the line of sight. As shown in the left panel of Fig \ref{fig:pca-scores}, PC1 is positively correlated with the total gas column density. The analysis of \cite{2017Gratier,2021Gratier} has shown that this first component provides an independent estimation of the total gas column density, expressed as a weighted average of the lines intensities. Using a random forest method, \cite{2021Gratier} have shown how different lines, the $(1-0)$ transition for \CO, \thCO, \CeiO, \HCOp and \NtwHp, can be combined to provide an accurate estimation of the gas column density across a broad range of values, from diffuse regions up to dense cores. The analysis of theoretical model predictions of molecular line intensities by \cite{2024Einig} using the mutual information between a physical parameter and the intensity of different lines with realistic S/N ratios shows that the total extinction (which is proportional to the total gas column density) is indeed strongly related to the intensity of selected lines and that the line selection depends on the extinction regime to be probed.

\subsection{PC2: core chemistry and role of the environment}\label{sec:PC2} 

The loadings of the studied lines for the second component, PC2, exhibit a wide range of variation with positive values for \CO, \twCN, \HCOp\ and \CCH\ among other lines, and negative values for CH$_3$OH, DCO$^+$, N$_2$H$^+$, DNC, \HthCOp amongst others (see Fig. \ref{fig:pca-loadings}, middle panel). This selection of lines implies that this component summarises the differences between dense, cold and shielded cores and those exposed to higher FUV radiation field. As shown by detailed chemical models deuterium fractionation is most efficient in cold and shielded gas where the ionization fraction is low \citep{2009Goicoechea, 2015Roueff}. By contrast, the CN and CCH radicals are efficiently produced in the outer layers of photodissociation regions, exposed to FUV radiation with a higher ionization fraction \citep{2005Pety,2017Pety,2009Goicoechea,2025Beslic}.

The middle panel of Figure \ref{fig:pca-scores} shows the variation of the PC2 scores for each core in the sample as a function of the \Gnotn\ parameter, which is a proxy for investigating the role of FUV radiation on the gas chemistry and ionisation fraction \citep{2025Beslic, 1997HollenbachTielens}. Low \Gnotn\ values ($<0.01$) correspond to dense and shielded regions where the dust is cool, while the larger values ($>0.1$) correspond to regions impacted by FUV radiation either because of a more intense radiation or because of relatively low gas densities and column densities which allow an easy penetration of the energetic UV photons. The strong relation between PC2 and \Gnotn\ (characterised by Pearson's correlation coefficient of $\approx0.8$) indicates that the observed variations of the line emission pattern from core to core are clearly related to the core environment conditions and that the FUV illumination is a main effect in determining the core chemical content. Indeed, as shown in Fig \ref{fig:pca-maps}, cores with positive PC2 scores are located in the vicinity of the NGC~2024 region and along the western ridge illuminated by $\sigma$Ori.  
The unbiased selection of cores using dust continuum emission allows to include both "classical" cold and dense cores with bright \HthCOp, \NtwHp, and \DCOp\ emission, but also a different population of cores exposed to FUV radiation with warmer dust temperatures and intense \twCN, \HCN\ and \HCOp\ emission. The first population is more numerous in our sample and has properties similar to those of well studied cold and dense cores in the Taurus or Perseus molecular clouds (e.g., L1544, \citealp{2023Jensen}). The column densities and densities encountered in the second core population, the warm cores with positive PC2 scores, largely overlap with those of the cold cores, though they do not extend to the most extreme values of the H$_2$ density and column density. These cores are mainly located in FUV exposed regions, at the boundaries of the NGC~2024 HII region, and at the dense cloud edge facing $\sigma Ori$. They have a distinct chemistry associated with higher dust temperatures, but otherwise similar masses and star forming capabilities. Identification of prestellar and protostellar cores based on molecular lines should therefore include enough tracers to find cores in a wide range of environments.

The three sources in NGC~2024, one of them identified as an outlier in this PCA analysis, have warm dust temperatures and negative PC2 values. Their position in the PC2 vs \Gnotn\ diagram clearly departs from the main trend in the sense that other cores with similar \Gnotn\ values or similar dust temperatures have a positive PC2 projection. These three cores have higher column densities, densities and masses than most other cores of the sample. Being associated with a massive star formation region, they exhibit a mixed emission pattern between the cold and the warm cores, with  intense \NtwHp, \HthCOp\ and CN emission, but a deficit in \DCOp.  

\subsection{PC3: sensitivity to gas density and freeze-out}\label{sec:PC3}

The loadings of the molecular lines for the third principal component present an interesting pattern. The relatively faint CO isotopologues \CeiO, \CseO\ and \thCO\ exhibit positive loadings while the loadings for \NtwHp, \CCH, \HNC\ and \DNC\ are negative (see Fig.\ref{fig:pca-loadings} right panel). Following \cite{2017Gratier}, we explored the relation between the mean gas density and PC3 as the gas density is one of the key parameter for controlling the line emission, together with the species abundance. As shown in the right panel of Fig. \ref{fig:pca-scores}, PC3 increases with increasing mean density up to $n \sim 10^{3.5}$ cm$^{-3}$, and then decreases again. Cores with higher mean densities than this threshold present a large scatter in their PC3 scores. The increase of PC3 score with increasing mean density can be understood as an excitation effect for the optically thin ground state transition of the \CeiO \, and \CseO \, isotopologues which have a moderate critical density of $\sim 10^{3.3}$ cm$^{-3}$ \citep{2024Roueff}. A higher density along the line of sight implies a more efficient collisional excitation for the same amount of molecules and a brighter line intensity. The effect of the mean density breaks down for densities larger than $\sim 10^{3.5}$ cm$^{-3}$ since the critical density for collisional excitation of the CO isotolopologue lines is reached.

The most negative PC3 scores are found for the high density cores. These objects are located in the highest density and column density regions, where the dust is cool (B9 and the dense ridge next to NGC~2023) as illustrated in Fig. \ref{fig:pca-maps}. They exhibit emission from deuterated species (DCO$^+$, DNC) and a deficit in \CeiO \, and \CseO \, emission due to freeze-out. Indeed, their dust temperatures are relatively low, consistent with the CO sublimation temperature of $\sim20$~K \citep{2022Minissale, 2005Oberg}. 
Given the relatively high densities along the line of sight, the freezing time scale for collisions between gas and cold grains, $t_{fr} = \frac{10^9}{n}$ yr \citep{1987Tielens,2015Boogert}, is lower than 10$^5$ years, the core lifetime. Freezing can therefore efficiently decrease the gas phase abundance of CO and its isotopologues, and contribute to the chemical enrichment of the ice mantles. Cores where freeze-out is efficient therefore display a deficit in CO isotopologue emission together with an enhanced emission for deuterated species and \NtwHp. The pattern we identify in PC3 is consistent with well studied cases in nearby molecular clouds where extreme freeze-out has been observed \citep{2024Pagani}.

\begin{figure}[h!]
    \centering
    \includegraphics[width=1\linewidth]{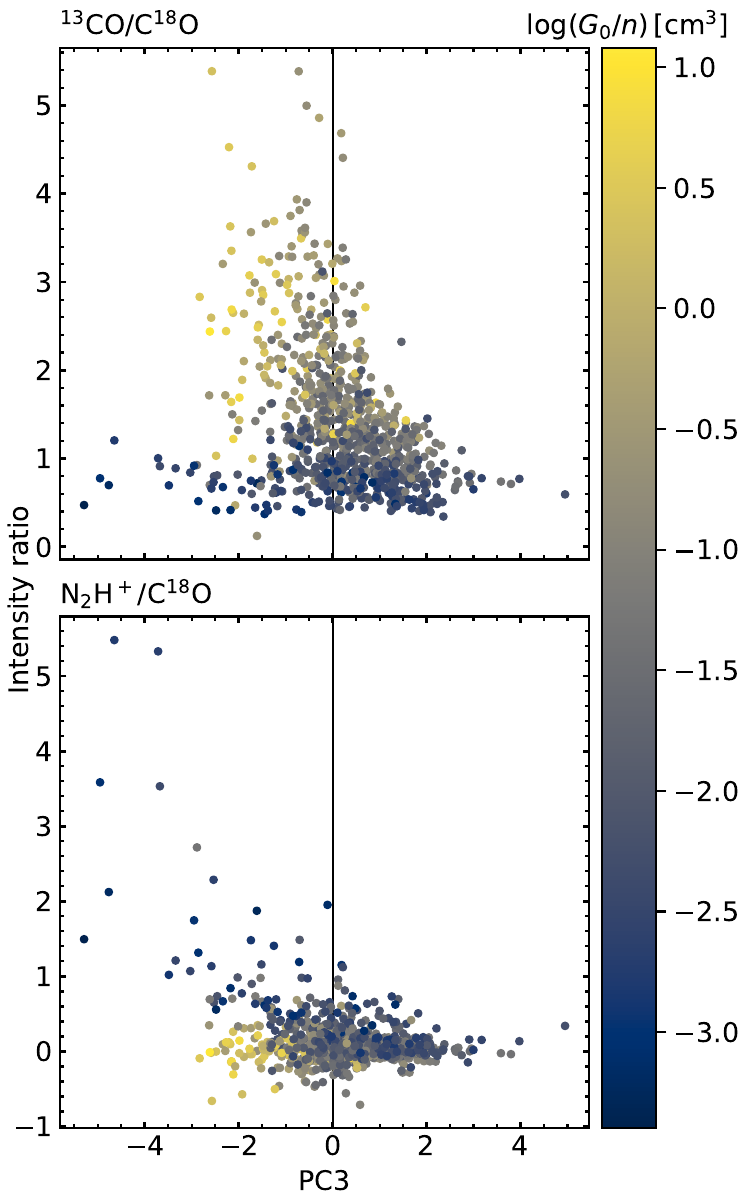}
    \caption{Intensity ratio of $\rm ^{13}CO/C^{18}O$ and $\rm N_2H^+/C^{18}O$ as a function of the PC3 scores. The colour of the points corresponds to $\rm log_{10} (G_0/n)\,[cm^{3}]$. Only the samples for which \CeiO\,is detected are shown here (\#878).}
    \label{fig:pc3-ratios}
\end{figure}

\subsection{Summary of the PCA analysis}
The first three PCs represent 64\% of the total variance of sample of 1003 cores and protostars traced by 25 molecular species. These three PCs are the most robust with respect to the presence of outliers in the sample. The physical interpretation of the meaning of these PCs allows us to identify the main causes of the variance. In addition to the total quantity of material along the line of sight, we have seen that the core environment quantified by the \Gnotn\ parameter and the mean core density $n$ are key parameters for explaining the core emission pattern. The third component quantifies deficit/excess of emission with respect to the mean value. Besides the density, it is also sensitive to the molecular depletion pattern caused by freeze-out. Through the means of the PC components, in particular the second and third, objects of unique line patterns are identified pointing to an ensemble of objects typically reported in literature as gravitationally collapsing prestellar cores and a secondary, more unique group of objects, exposed to higher radiation field traced by high \Gnotn, clearly affecting their line emission pattern and thus chemical composition. Having a diversity of emission lines is important to fully account for the diversity of conditions. Besides the classical "cold dense core" tracers \NtwHp\ and \HthCOp, information on CO isotopologues and on UV sensitive species like CN is useful to obtain a full and unbiased characterisation of cores of all types.

\section{Discussion}
\label{sec:discussion}

\subsection{Robustness of cores}

Studies of the continuum emission and subsequently core mass function (CMF) along with infrared observation are excellent in identification of starless and protostellar cores, but are subject to greater uncertainty regarding the transition to prestellar, gravitationally unstable cores due to lack of dynamical and chemical composition insights, which provide information on external effects (turbulence, external pressure, heating) and gas processing on case-to-case basis. In this work, we proposed to complete the picture by statistically examining the molecular emission.

Through the analysis of the velocity dispersion of cores, we demonstrated that the Bonnor-Ebert criterium while assuming a single internal core temperature does not suffice in the inference of robustness of cores. An estimate of the temperature or dispersion is needed. Additionally, what is also important to emphasise is that cores in Orion B, particularly those exposed to FUV and consequently experiencing stronger turbulence are far from being close to Bonnor-Ebert spheres and need additional consideration of external pressure and internal motions \citep{2025Moon}.

PCA analysis performed on cores shown here, and in particular the second PC, revealed that chemical diversity of cores is driven by the \Gnotn\ parameter. It proves the importance of this particular parameter for the purpose of identifying prestellar cores along the Bonor-Ebert criterium, as it helps in accounting for the environmental impact and internal core chemistry. Such a parameter is also easily accessible when infrared observations are at hand (like those of \textit{Herschel}) as $G_0$ can be calculated directly and density is retrievable from \NH column densities. Then, core identification studies that use chemical tracers such as \NtwHp and $\rm H_2D^+$ lead to a biased sample of cold and dense prestellar cores. Here, we show that by using complementary tracers such as \twCN, \HCN, \CCH\ or \CS\ could be helpful in completing the sample to include cores that are exposed to a higher \Gnotn. Starless and candidate prestellar cores follow general properties described by the first 3 PCs affirming their classification done by K20, whereas the PC projections for the robust and protostellar objects deviate more indicating more diverse properties (see Fig.~\ref{fig:pc_pc}). Most robust cores have projections similar to the rest of the sample, and a subset of them have more extreme PC projections and closer to those of protostellar objects. The only information that can be deduced from this is that these objects exhibit a stronger than average variance compared to the rest of the sample, which could be caused by their unique chemical content (possibly due to a more evolved status) or different environmental influences compared to the rest of the sample. The statistical analysis shown in this work is helpful in identifying such cases, but surely does not answer the question of evolutionary status. The only reliable mean of accessing such information is through detailed studies of the core structure, which firstly and most importantly, disentangle the gas structure along the line of sight, something which has been done for example by \citet{2024Segal} for the Horsehead and, ideally, are followed by higher resolution observations good enough to retrieve the physical scales of the cores. 

\begin{figure*}
    \centering
    \includegraphics[width=1\linewidth, trim=0 5 0 0, clip]{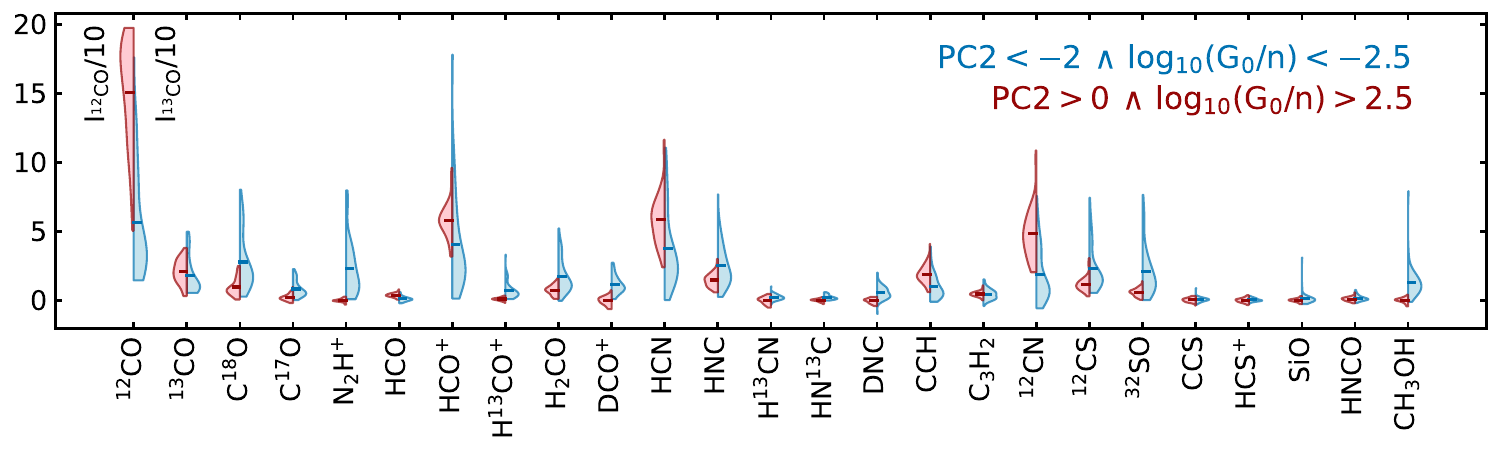}
    \caption{Comparison of probability distribution of the analysed lines for two sub-samples of cores. The two sub-samples of cores are defined by two boxes:  [PC2 scores $\leq-2.5$ and $G_0/n \leq -2$] and [PC2 scores $\geq 2.5$ and $G_0/n \geq 0$]. The fluxes of \CO \, and \thCO\, have been divided by 10 to ease the comparison with other lines. The horizontal lines over-plotted on each of the half-violin distribution plots corresponds to the median flux in the subsample.}
    \label{fig:box-flux}
\end{figure*}

\subsection{Environmental impact on the molecular emission}

In Section \ref{sec:PC2} we have shown an interpretation of the second principal component, which traces differences in the core properties that can be explained with the \Gnotn\ parameter. By taking the most extreme regimes of the two parameters (shown in Fig. \ref{fig:pca-scores}) defined by two boxes: [$\rm PC2 > 0\, \wedge \, log_{10}(G_0/n) > 2.5$] and [$\rm PC2 < -2\, \wedge \, log_{10} (G_0/n) < -2.5$], both of which contain 50 objects, and examining their flux probability distributions (Fig. \ref{fig:box-flux}, it shows more quantitatively what was revealed by the loadings of the second component. The first group of objects resides exclusively around the NGC~2024 region (dark purple points in Fig. \ref{fig:pca-maps}b) and it comprises starless (\#22) or prestellar cores (\#28). The latter group resides mostly in the NGC~2023 region, B9 region with a few objects in the Cloak and the Flame, and consists of mostly robust prestellar cores (\#39) and protostars (\#9). These groups mark two sub-categories of cores that are exposed to the most extreme physical conditions and thus vary significantly in their chemical markers, forming history and evolutionary stage. Most of the starless core and prestellar core studies focus on the second sub-category of cores, but results shown in this work share new insights and underline the need for further studies of the cores which are exposed to more extreme, FUV exposed environments. 

Then in section \ref{sec:PC3} we presented the interpretation of the third component whose variation can be traced by the density $n$. While it was shown in \ref{sec:PC3} that at densities $n>10^{3.5}$cm$^{-3}$ the phenomenon of freeze-out is observed, a secondary, environmental effect of dissociation can be identified too. In Fig. \ref{fig:pc3-ratios}, the intensity ratios of \thCO\ and \NtwHp\ w.r.t. \CeiO\ are shown. For PC3 $<-3$ a much enhanced $\rm N_2H^+/C^{18}O \gtrsim 3$ is observed, expected for the core freeze-out \citep{2013Lippok}. The most extreme cases correspond to cores and protostars in the B9 region (core 781, protostar 993-1 and 993-2, and a protostar 995). Then, a unique group of objects where $\rm -1>PC3>-3$ and $log_{10} (G_0/n) > 0$ show an enhancement of $\rm ^{13}CO/C^{18}O$. This ratio is expected to be enhanced due to fractionation of \thCO\ and photo-dissociation of \CeiO\ \citep{2021Roueff}, which is likely caused through exposure of cores to the UV radiation from NGC~2024 HII region. What is expected for these objects are different triggering and sequential formation mechanisms as compared to typical cold cores. 
It has been shown that the first generation of OB stars of the NGC~2024 region was likely triggered by a cloud-cloud collision \citep{2017Ohama}, with the merge point coinciding with the highest column density (\NH > $10^{23}\,[\rm cm^{-2}]$). This location hosts at least two pre-stellar cores (646, 667) and an active protostar (988), which our analysis shown to be extremely unique with respect to the rest of the cores in the cloud, whose formation could had been triggered by the merger. Subsequently, cores surrounding the NGC~2024 HII region could be a secondary generation of cores being triggered by the feedback of the young OB stars illuminating the region. It has been shown that cores residing in the pillars of the Eagle Nebula are likely undergoing linear sequential star formation due to triggered star formation caused by the irradiation from the M16 cluster \citep{2002Fukuda}. The effectiveness of such processing was further demonstrated with direct observations of hundreds of protostars at the edges of the pillars \citep{2025Wen}. Similar triggered and sequential star formation could be observed in the illuminated region of NGC~2024. 

\section{Conclusions}\label{sec:conclusion}

In this article, we presented observations of the Orion B giant molecular cloud in the framework of the ORION B mapping programme targeting 25 molecular tracers, some of which were not published before. The primary objective of this work was to study the diversity of prestellar cores and protostars residing within this GMC. The sample of cores was selected from the catalogue extracted based on the \textit{Herschel} observations by \citet{2020Konyves}. 
To combine information of all the molecular lines, we applied the Principal Component Analysis to a sample of 1003 extracted core fluxes.
We found that the first PC traces the global behaviour of the increase of line emission with the bulk gas or total column density and is therefore directly correlated with \NH. The second component highlights the differences in the chemistry between populations of cores that are shielded and exposed to FUV radiation and is directly traceable with the \Gnotn\ parameter. The third component describes a combined effect of the increase of molecular abundance of optically thin species with the mean density up until $n>10^{3.5}$~cm$^{-3}$, but also gives insights into the effect of fractionation and freeze-out for cores with highest densities. Additional PC analysis performed on the subsamples of cores, or otherwise LOO PCA, served to determine the robustness of our PCA analysis, but also revealed an effective method of detecting strong outlier sources, which present unique core emission patterns as compared to the rest of the cores in the cloud. Additional to the PCA, we used the high spectral resolution observations of \CeiO$(1-0)$ to investigate the core line widths and their dynamical status. Having measured that on average the line widths of cores in Orion B are larger than for other star forming sites, we argued that the criterium of core robustness, repeatedly referenced in previous continuum and infrared prestellar core studies, may be strongly biased due to the lack of dynamical information. We also showed that the \CeiO \, and \CseO$(1-0)$ lines trace the same environment with a fixed line intensity ratio of \CseO/\CeiO=$0.292\pm 0.001$.

\section{Data availability}
\label{sec:data}
The first part of the ORION-B data is available at the IRAM Data Management System (IRAM-DMS) accessible through the link \href{https://oms.iram.fr/?dms=showprograms&pageId=7}{https://oms.iram.fr/oms.iram.fr/large-programs} [Project: ORION-B (Outstanding Radio-Imaging of OrioN-B)]. The archive will be enriched by the data cubes covering the five square degree field of view by the end of 2026. Observations which have not been published before include data cubes centred on the lines from \DCOp, \DNC, \HNCO, \HNthC, \HCSp\ and \CCS\ listed in Table \ref{tab:molecular-species}. Table \ref{tab:cores} and an additional table containing core intensity measurements with their estimated uncertainties are only available in electronic form at the CDS via anonymous ftp to cdsarc.u-strasbg.fr (130.79.128.5) or via http://cdsweb.u-strasbg.fr/cgi-bin/qcat?J/A+A/. The full set of measurements obtained in the PCA analysis along with other measurements presented in this work can be made available upon request to the lead author (Helena J. Mazurek, helena.mazurek@obspm.fr).

\begin{acknowledgements}
This work is based on observations carried out under project numbers 019-13, 022-14, 145-14, 122-15, 018-16, and finally the large program number 124-16 with the IRAM 30m telescope. IRAM is supported by INSU/CNRS (France), MPG (Germany) and IGN (Spain). This research has also made use of data from the \textit{Herschel} Gould Belt Survey (HGBS) project (http:// gouldbelt-herschel.cea.fr). This work received support from the French Agence Nationale de la Recherche through the DAOISM grant ANR-21-CE31- 0010, and from the Programme National “Physique et Chimie du Milieu Inter-stellaire” (PCMI) of CNRS/INSU with INC/INP, co-funded by CEA and CNES. JRG and MGSM thank the Spanish MCINN for funding support under grant PID2023-146667NB-I00 funded by MCIN/AEI/10.13039/501100011033. Part of this research was carried out at the Jet Propulsion Laboratory, California Institute of Technology, under a contract with the National Aeronautics and Space Administration (80NM0018D0004). D.C.L. acknowledges financial support from the National Aeronautics and Space Administration (NASA) Astrophysics Data Analysis Program (ADAP).
\end{acknowledgements}

%
%

\bibliographystyle{aa} %
\bibliography{bibliography}

\begin{appendix}

\section{Core Sample}\label{app:cores}

Table \ref{tab:cores} presents our sample. In this work, we use the core positions reported by \cite{2020Konyves}. For a small fraction of objects, there are noticeable offsets between the positions and the peak of the \NH\ column density map. These often follow closely emission of \NtwHp, which is a known tracer of high density and low temperature, and thus also of the more evolved pre-stellar cores. By comparing the catalogue positions and an SNR clipped moment 0 map of \NtwHp we identified about nine objects where an offset is evident, that includes cores 453, 546, 547, 618, 652, 792, 829, 987 and 996. From the moment 0 map of \NtwHp, we measured that the flux integrated over all cores and their radii defined by K20 accounts for $\sim85\%$ of the total flux in the map. We identify, however, that regions of significant ($\gtrsim 10^{23}\,[\rm cm^{-2}]$) \NH\ column densities such as NGC~2024 and NGC~2023 have significant extended emission missed. This could point to a caveat of the continuum extraction algorithm faltering for regions where more extended emission is observed.

\begin{figure}[ht!]
    \centering
    \includegraphics[width=1\linewidth]{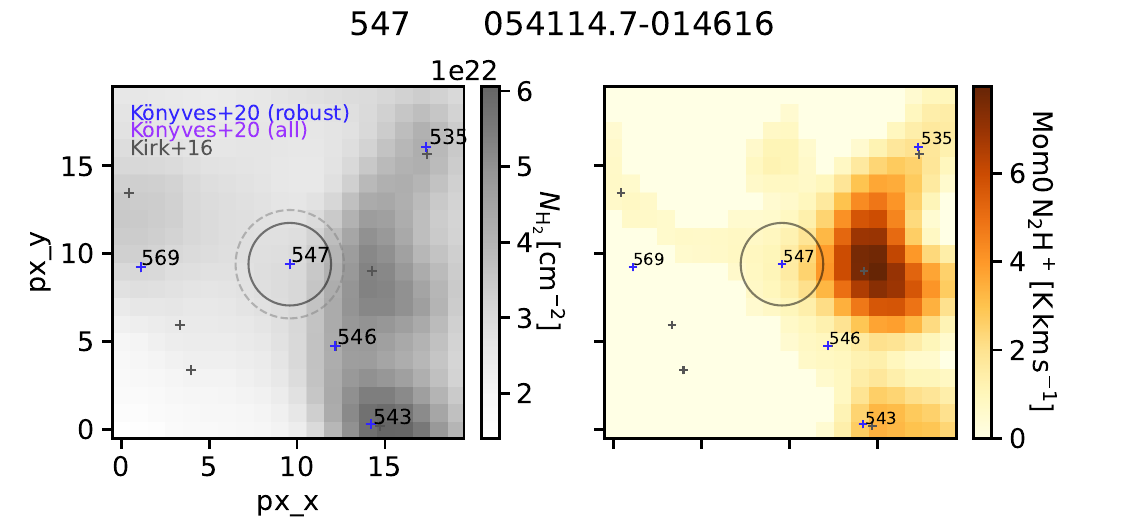}\\
    \includegraphics[width=1\linewidth]{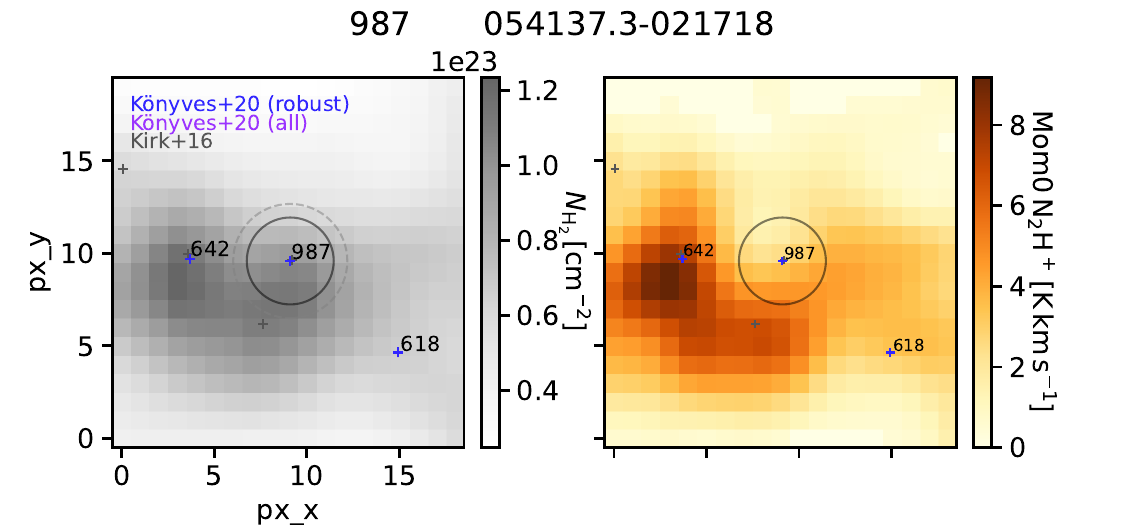}
    \caption{Examples of cores, 547 and 987, affected by position discrepancies w.r.t. \NtwHp and \NH column density. Positions reported by \cite{2020Konyves} and \cite{2016Kirk} are annotated for reference in dark purple (with robust cores in dark blue) and gray, respectively. The gray circle drawn in the column density map shows the core size and the dashed circle shows the expected core size at the angular resolution of the emission line data. The gray circles on the \NtwHp\ maps show the core size.}
    \label{fig:core-offsets}
\end{figure}

Moreover, the positions of the offset affected cores are often better defined in the catalogue reported by \cite{2016Kirk}. Two examples of such cases are shown in Fig. \ref{fig:core-offsets}, which illustrate the issue of missing the most prominent \NtwHp\ emission around these cores. This discrepancy can be explained by the general property of the molecular emission being more closely coupled with the colder dust observable at longer wavelengths $\gtrsim 300 \mu m$ \citep{2026Pineda, 2017Friesen}. Due to this property and the fact that the core extraction algorithm is optimised on multiple wavelength between 70-500 $\mu m$ (with a higher weight adopted for shorter wavelengths on account of their higher resolution \citep{2012Menshchikov}), the position accuracy w.r.t. the molecular emission of cores is to be affected. However, by verifying the sample of objects with resolvable \NtwHp\, emission and considering the resolution of the observations used in this work, we can confidently report that these discrepancies should not have a significant affect on the methodology and conclusions presented herein.

\begin{table*}[h]
\footnotesize
\centering
\rotatebox{90}{%
\begin{minipage}{0.85\textheight}
\caption{Summary of the sample of cores and protostars studied in this work$^a$}
\label{tab:cores}
\begin{tabular}{l|l|l|l|l|l|l|l|l|l|l}
\hline\hline
Index & \makecell{K20\\rNO} & \makecell{Core\\type} & $\rm Name\, {HGBS\_J*}$ & $\rm RA_{2000}$ & $\rm Dec_{2000}$ & $\rm R_{core}$ [pc] & $N_{\rm H_{2}} [\rm cm^{-2}]$ & $T_{\rm dust}$ [K] & $G_0$ [Habing] & $n \rm \,[cm^{-3}]$ \\ 
\hline
\multicolumn{11}{c}{\textbf{Starless cores}} \\ \hline
1 & 5 & 1 & 053919.1-012357 & 5:39:19.17 & -1:23:57.90 & 0.060 & 3.74e+21 & 20.61 & 14.86 & 125.51 \\ 
2 & 7 & 1 & 053925.2-012443 & 5:39:25.22 & -1:24:43.10 & 0.058 & 5.25e+21 & 19.79 & 16.07 & 677.83 \\ 
3 & 8 & 1 & 053927.4-012541 & 5:39:27.47 & -1:25:41.10 & 0.053 & 4.69e+21 & 20.05 & 15.20 & 359.60 \\ 
4 & 9 & 1 & 053937.3-013027 & 5:39:37.36 & -1:30:27.50 & 0.048 & 4.21e+21 & 20.30 & 13.93 & 283.55 \\ 
5 & 10 & 1 & 053939.7-012759 & 5:39:39.77 & -1:27:59.40 & 0.042 & 5.01e+21 & 21.59 & 23.73 & 443.87 \\ 
6 & 13 & 1 & 053956.9-012854 & 5:39:56.94 & -1:28:54.00 & 0.044 & 3.32e+21 & 23.68 & 25.16 & 79.97 \\ 
7 & 15 & 1 & 054009.3-012759 & 5:40:09.38 & -1:27:59.00 & 0.058 & 3.11e+21 & 24.57 & 28.28 & 65.03 \\ 
8 & 16 & 1 & 054011.5-013920 & 5:40:11.55 & -1:39:20.70 & 0.064 & 4.16e+21 & 24.41 & 33.99 & 81.96 \\ 
9 & 17 & 1 & 054013.1-013026 & 5:40:13.13 & -1:30:26.50 & 0.035 & 4.60e+21 & 24.22 & 37.01 & 818.72 \\ 
10 & 18 & 1 & 054013.8-013000 & 5:40:13.83 & -1:30:00.80 & 0.044 & 4.68e+21 & 24.09 & 36.67 & 886.80 \\ \hline
\multicolumn{11}{c}{\textbf{Prestellar cores}} \\ \hline
501 & 31 & 2 & 054038.4-014315 & 5:40:38.41 & -1:43:15.60 & 0.084 & 8.46e+21 & 26.60 & 105.85 & 220.58 \\ 
502 & 34 & 2 & 054041.0-015627 & 5:40:41.07 & -1:56:27.90 & 0.035 & 5.97e+21 & 39.46 & 632.98 & 233.74 \\ 
503 & 36 & 2 & 054043.7-015453 & 5:40:43.70 & -1:54:53.60 & 0.047 & 5.49e+21 & 43.26 & 917.09 & 140.19 \\ 
504 & 37 & 2 & 054047.2-015325 & 5:40:47.22 & -1:53:25.80 & 0.046 & 6.26e+21 & 41.43 & 839.95 & 142.73 \\ 
505 & 38 & 2 & 054047.9-013822 & 5:40:47.94 & -1:38:22.30 & 0.046 & 1.03e+22 & 24.44 & 79.50 & 474.63 \\ 
506 & 39 & 2 & 054048.1-020739 & 5:40:48.14 & -2:07:39.90 & 0.062 & 7.38e+21 & 26.45 & 89.20 & 240.17 \\ 
507 & 46 & 2 & 054053.4-013821 & 5:40:53.40 & -1:38:21.80 & 0.052 & 1.06e+22 & 23.44 & 64.23 & 657.47 \\ 
512 & 54 & 2 & 054055.4-021718 & 5:40:55.40 & -2:17:18.90 & 0.040 & 6.12e+21 & 27.93 & 102.49 & 130.86 \\ 
515 & 62 & 2 & 054058.2-022542 & 5:40:58.29 & -2:25:42.90 & 0.039 & 1.02e+22 & 20.83 & 34.07 & 296.41 \\ 
518 & 70 & 2 & 054059.6-015356 & 5:40:59.68 & -1:53:56.20 & 0.035 & 8.96e+21 & 39.89 & 990.17 & 356.29 \\ \hline
\multicolumn{11}{c}{\textbf{Robust cores}} \\ \hline
508 & 47 & 2 & 054053.4-022704 & 5:40:53.48 & -2:27:04.10 & 0.054 & 2.62e+22 & 19.19 & 52.41 & 12093.27 \\ 
509 & 49 & 2 & 054054.4-022757 & 5:40:54.49 & -2:27:57.20 & 0.094 & 2.17e+22 & 20.30 & 55.04 & 8960.84 \\ 
510 & 50 & 2 & 054054.8-015313 & 5:40:54.85 & -1:53:13.50 & 0.055 & 9.38e+21 & 41.33 & 1233.63 & 653.57 \\ 
511 & 52 & 2 & 054055.1-011307 & 5:40:55.16 & -1:13:07.50 & 0.120 & 1.50e+22 & 19.16 & 34.61 & 1816.33 \\ 
513 & 57 & 2 & 054057.1-014827 & 5:40:57.12 & -1:48:27.40 & 0.039 & 9.77e+21 & 37.59 & 781.93 & 674.63 \\ 
514 & 61 & 2 & 054058.0-020729 & 5:40:58.09 & -2:07:29.50 & 0.057 & 2.89e+22 & 20.29 & 79.80 & 13429.29 \\ 
516 & 64 & 2 & 054058.6-020842 & 5:40:58.66 & -2:08:42.10 & 0.046 & 5.11e+22 & 19.42 & 111.17 & 47559.45 \\ 
517 & 65 & 2 & 054058.8-022715 & 5:40:58.81 & -2:27:15.90 & 0.051 & 1.96e+22 & 18.45 & 33.47 & 2888.98 \\ 
519 & 72 & 2 & 054059.9-011256 & 5:40:59.90 & -1:12:56.10 & 0.074 & 1.73e+22 & 18.14 & 30.04 & 2795.80 \\ 
521 & 79 & 2 & 054101.6-012622 & 5:41:01.60 & -1:26:22.00 & 0.039 & 1.02e+22 & 21.74 & 42.18 & 1328.09 \\ \hline
\multicolumn{11}{c}{\textbf{Protostellar cores}} \\ \hline
982 & 14 & 3 & 054003.9-021647 & 5:40:03.95 & -2:16:47.50 & 0.038 & 2.77e+21 & 19.37 & 6.25 & 953.19 \\ 
983 & 193 & 3 & 054125.2-021807 & 5:41:25.21 & -2:18:07.70 & 0.038 & 1.64e+23 & 17.92 & 218.65 & 174917.36 \\ 
984 & 209 & 3 & 054127.5-014758 & 5:41:27.53 & -1:47:58.00 & 0.053 & 2.38e+22 & 36.98 & 1665.34 & 4651.90 \\ 
985 & 220 & 3 & 054129.7-022318 & 5:41:29.70 & -2:23:18.90 & 0.038 & 5.31e+22 & 17.79 & 72.06 & 18489.02 \\ 
986 & 222 & 3 & 054129.8-022117 & 5:41:29.88 & -2:21:17.60 & 0.035 & 9.87e+22 & 17.07 & 104.66 & 45444.16 \\ 
987 & 273 & 3 & 054137.3-021718 & 5:41:37.37 & -2:17:18.60 & 0.041 & 9.15e+22 & 22.76 & 476.79 & 41434.52 \\ 
988 & 320 & 3 & 054144.3-015445 & 5:41:44.37 & -1:54:45.40 & 0.040 & 1.86e+23 & 46.71 & 39383.81 & 191688.03 \\ 
989 & 428 & 3 & 054202.9-020745 & 5:42:02.93 & -2:07:45.20 & 0.037 & 6.44e+22 & 17.92 & 85.57 & 80756.05 \\ 
990 & 535 & 3 & 054227.8-012003 & 5:42:27.81 & -1:20:03.00 & 0.039 & 6.81e+22 & 17.00 & 68.72 & 35979.66 \\ 
991 & 593 & 3 & 054245.4-011616 & 5:42:45.43 & -1:16:16.10 & 0.039 & 8.63e+22 & 14.24 & 33.95 & 84791.93 \\ 
\hline\end{tabular}
\tablefoot{
\tablefoottext{a}{Only 10 entries of each of the core types (1 - starless, 2 - prestellar/robust, 3 - protostars) are listed here. The full catalogue is accessible online as described in section \ref{sec:data}. The first column contains the running index of cores, while the second column contains the core running index introduced by K20.}
}
\end{minipage}
}
\end{table*}

\section{Ratio between \CseO\, and \CeiO}

\begin{figure*}[hbt!]

    \includegraphics[width=0.5\linewidth, trim = 0 0 0 0, clip]{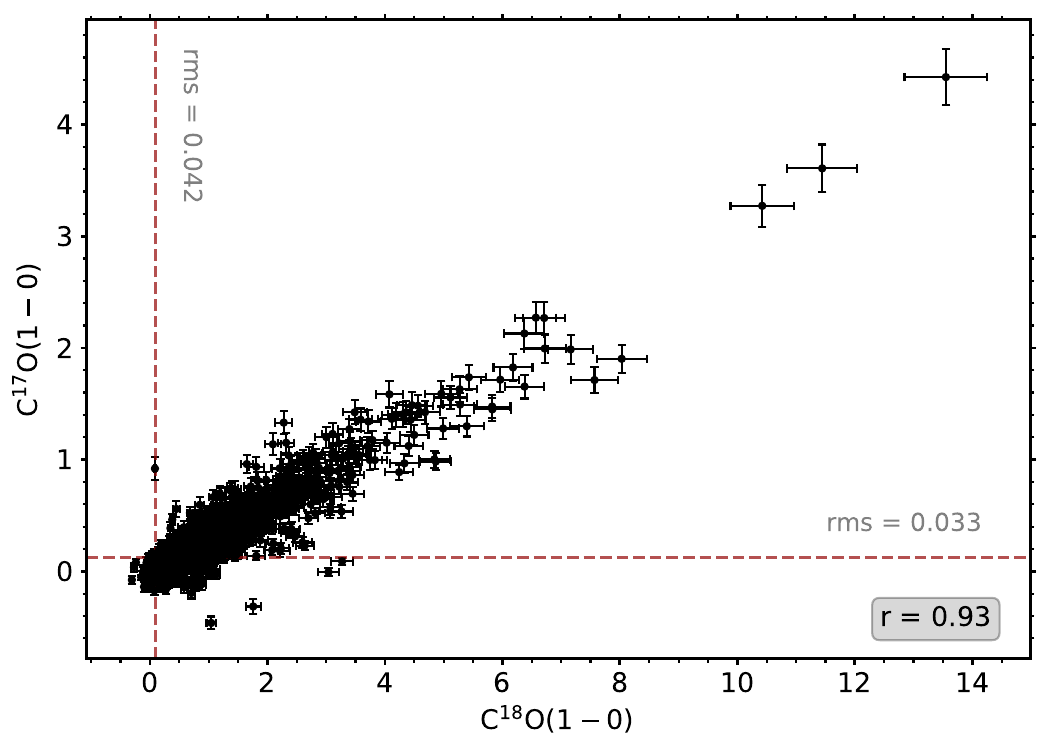}
    \includegraphics[width=0.5\linewidth, trim = 0 0 0 0, clip]{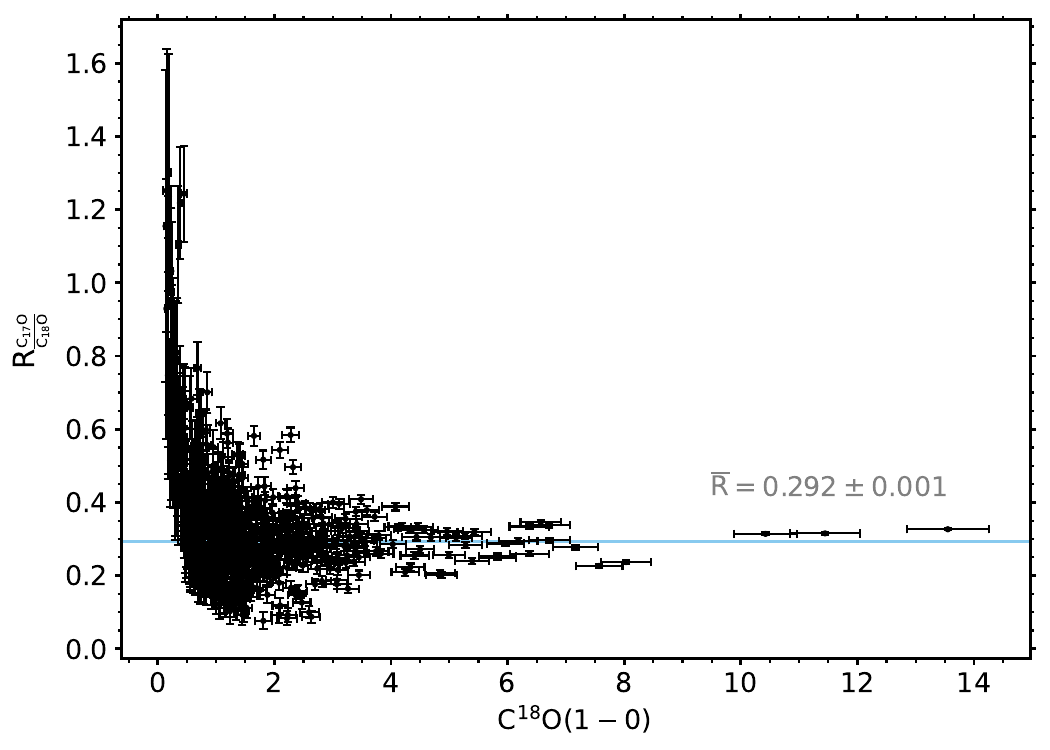}
    \caption{Comparison of the intensities of the \CseO\, and \CeiO\, $J=1-0$ lines. The left panel illustrates the excellent linear correlation of the lines and the right panel shows the uniformity of the \CseO/\CeiO \, ratio as a function of the \CeiO $(1-0)$ intensity.}
     \label{fig:co_iso_ratio}
\end{figure*}

The two isotopologues \CseO \, and \CeiO \,  are the two most correlated lines we observe for the cores with a Pearson's correlation coefficient of 0.93. After filtering out the points below the rms level of both molecules, their average ratio comes to $R = 0.292 \pm 0.001$ as shown in Fig.\ref{fig:co_iso_ratio}.

\section{Full results of the PCA analysis}
\label{pca:full}

This section presents additional information on the PCA analysis. Fig. \ref{fig:loading_app0} shows the full set of PC loadings for the whole sample (purple bars) and for the sample cleaned from outliers (orange bars), with their averages measured over all LOO trials shown as brighter bars. While the measured PC loadings and the averaged loadings are expected to be aligned and consistent between the different trials, some PCs show discrepancies (e.g., PC5 or PC17).
This is expected, particularly when the stronger outliers are removed, as their absence can shift the distribution of variance closer to the true axis of variation. The PC scores for all 25 PCs are shown in Fig.~\ref{fig:pc_pc}. 

\begin{figure*}
    \centering
    \includegraphics[width=\linewidth]{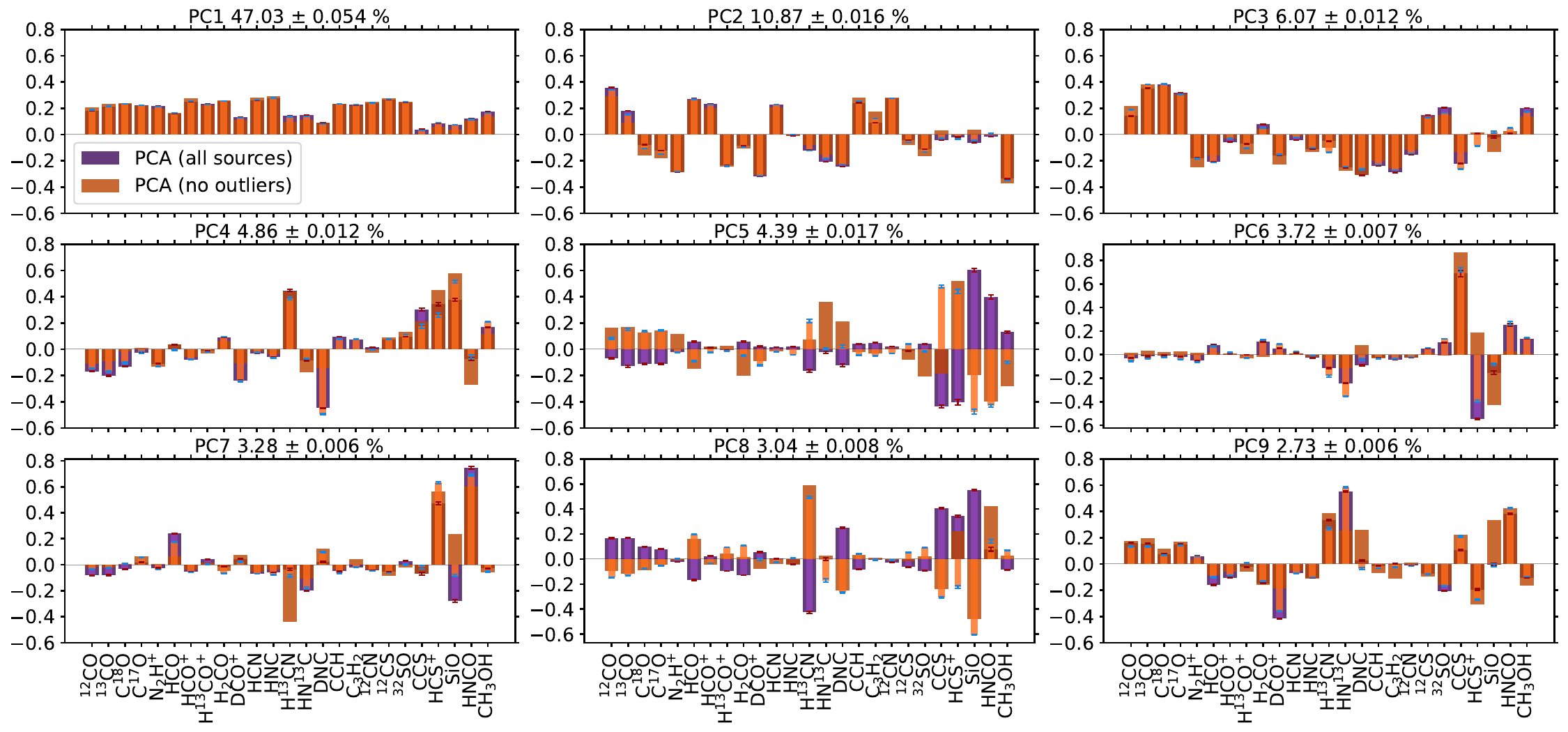}
    \caption{PC loadings for the analysis with and without outliers, in orange and purple respectively. The brighter bars indicate average loadings with the error bars measured as the standard deviation of the loadings over all LOO trials.}
     \label{fig:loading_app0}
\end{figure*}

\begin{figure*}[h]
    \addtocounter{figure}{-1}
    \centering
    \includegraphics[width=\linewidth]{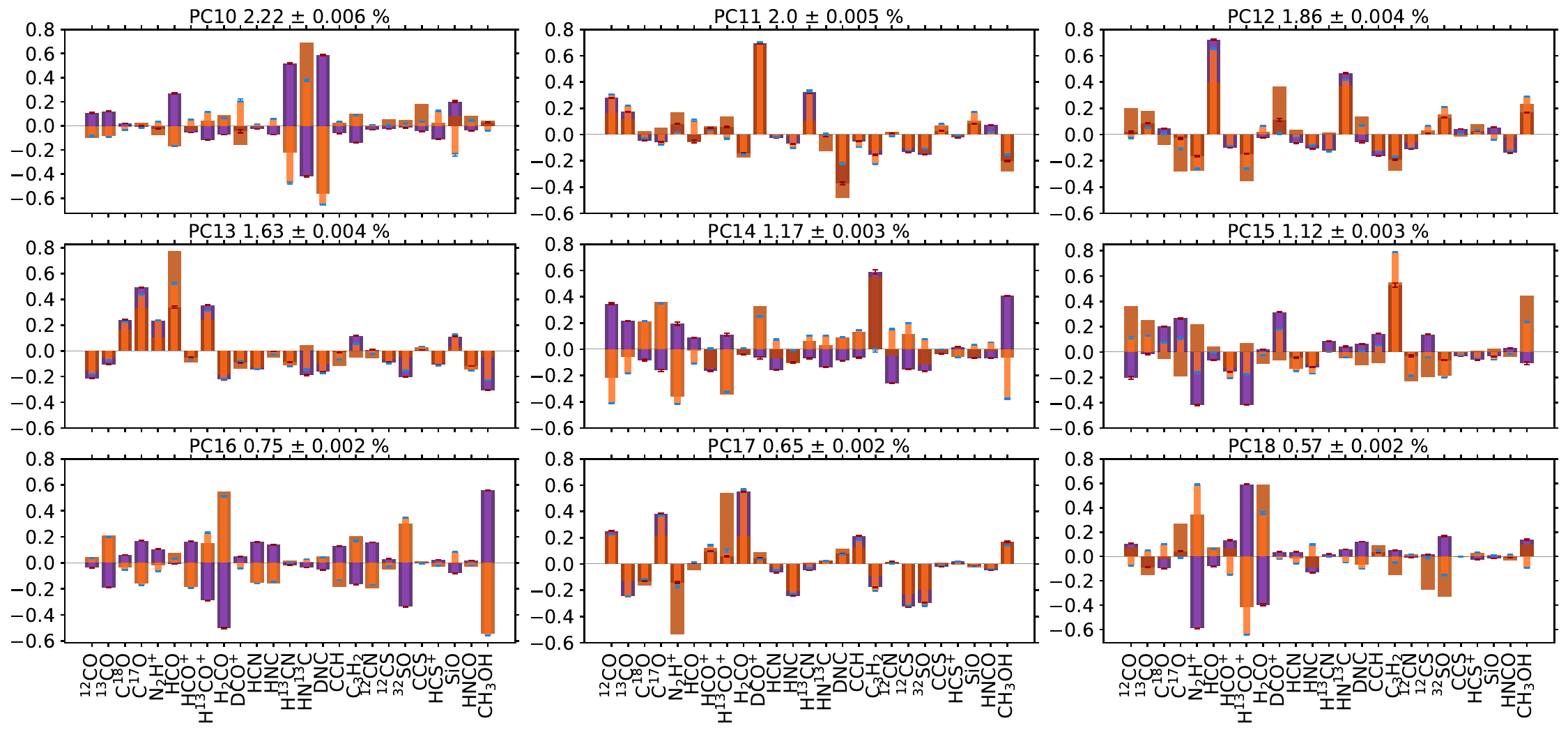}
    \caption{continued. PC loadings from 10 to 18.}
    \addtocounter{figure}{1}
    \label{fig:loading_app1}
\end{figure*}

\begin{figure*}[h]
    \addtocounter{figure}{-2}
    \centering
    \includegraphics[width=\linewidth]{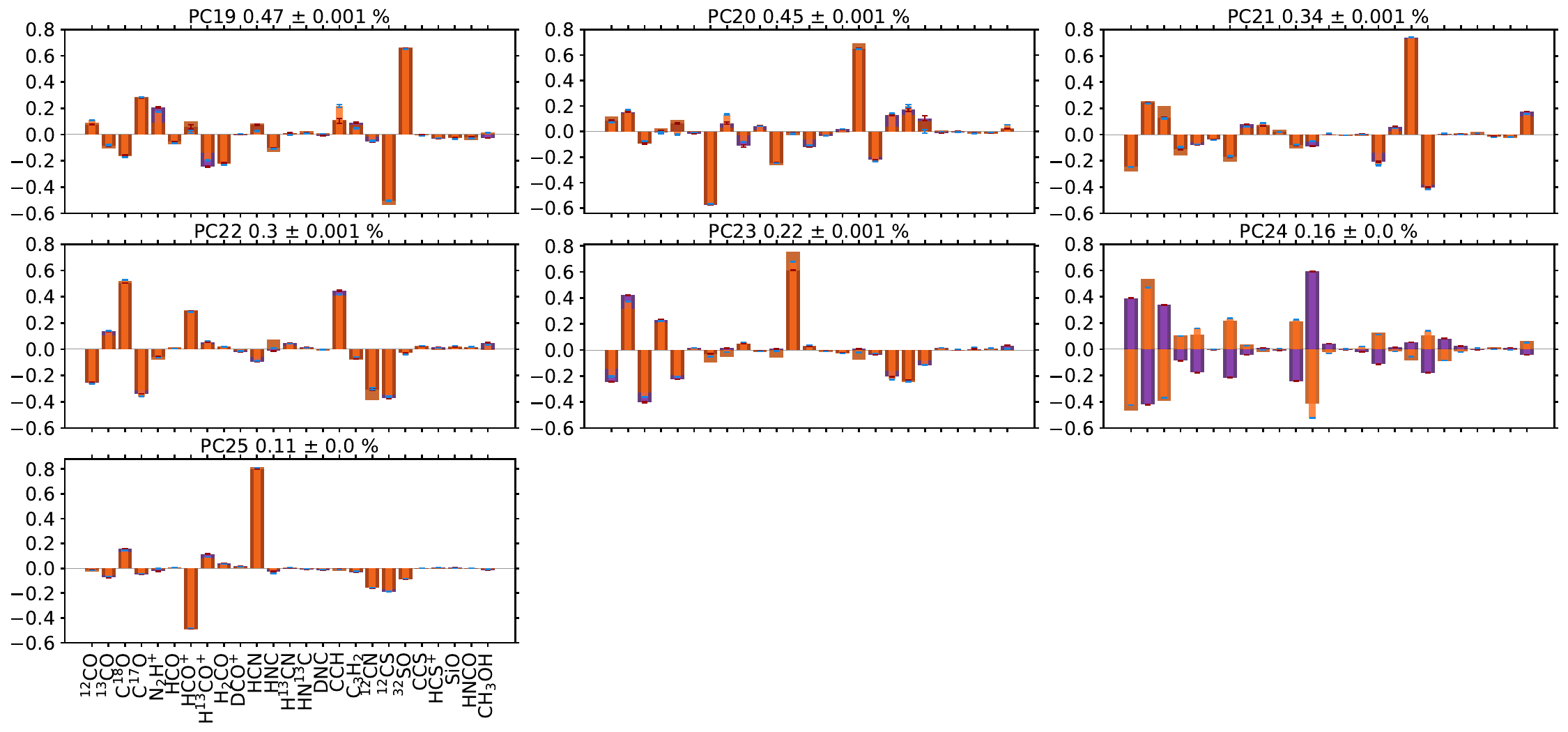}
    \caption{continued. PC loadings from 19 to 25}
    \addtocounter{figure}{1}
     \label{fig:loading_app2}
\end{figure*}

\begin{figure*}[h]
\addtocounter{figure}{-1}
    \centering
    \includegraphics[width=0.92\linewidth, trim = 0 10 0 10, clip]{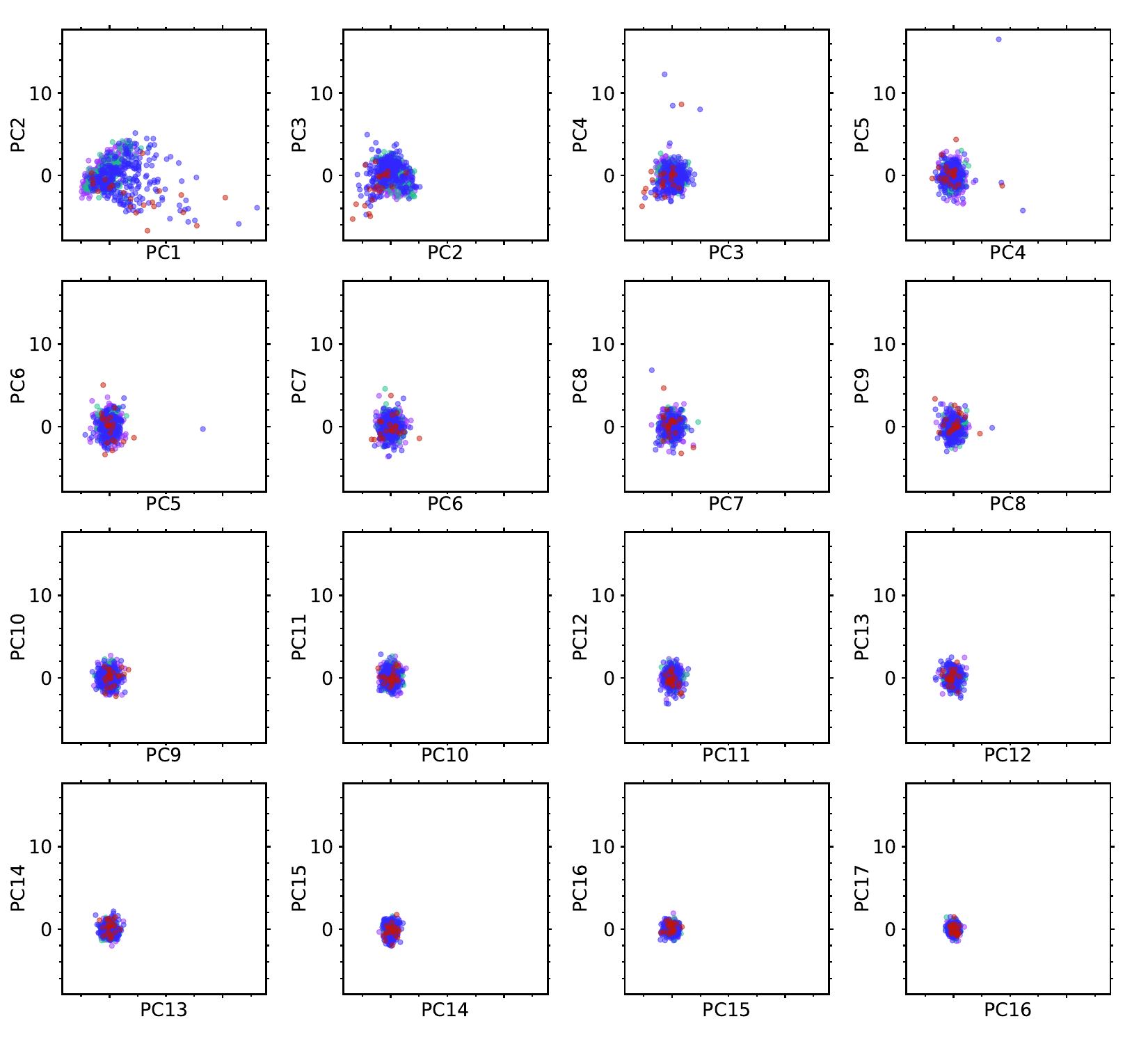}
    \includegraphics[width=0.92\linewidth, trim = 0 330 0 20, clip]{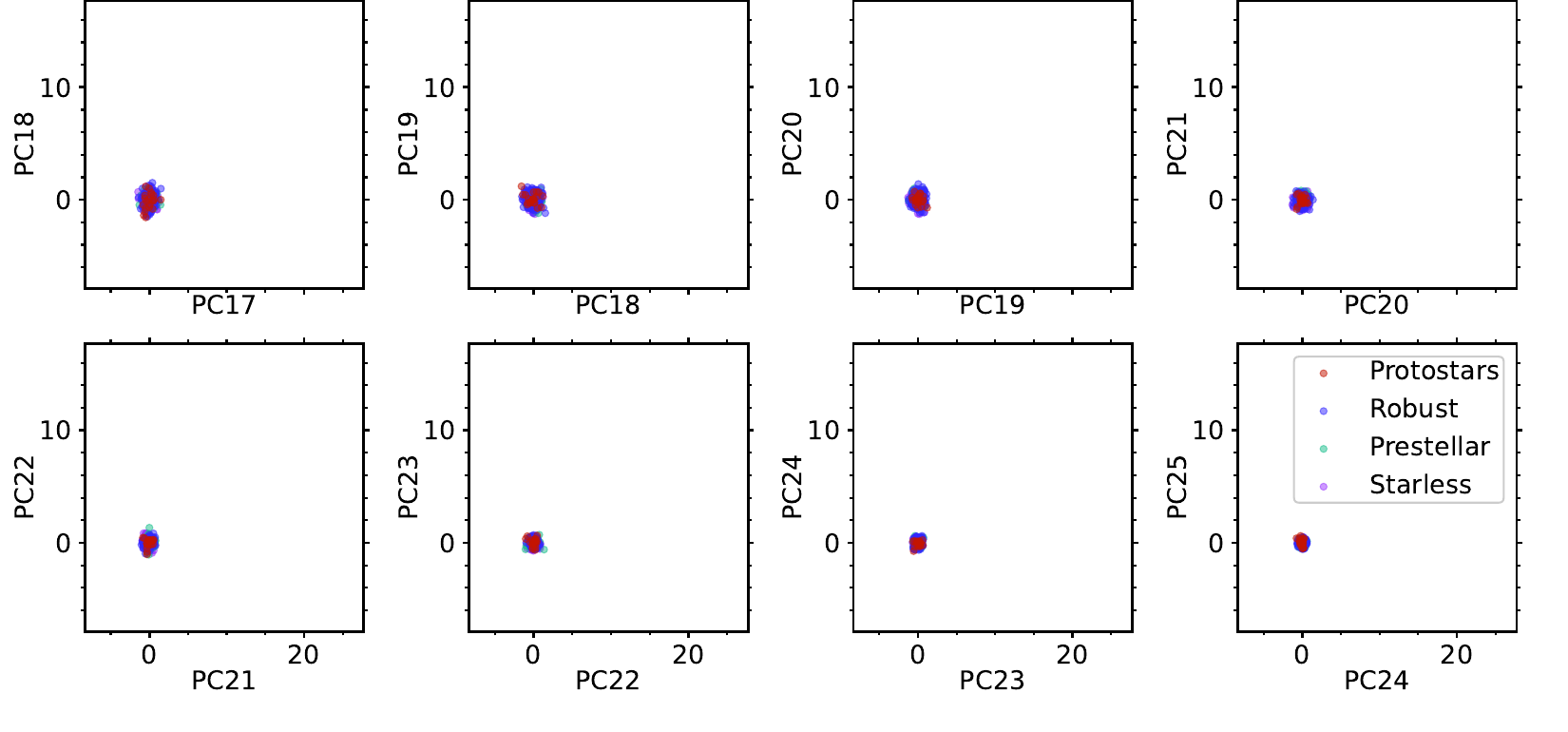}
    \caption{Presentation of the consecutive PC scores of the 1003 samples. Robust cores are shown in blue, prestellar cores in cyan and starless cores in magenta. The cores hosting protostars are shown in red.}
     \label{fig:pc_pc}
\end{figure*}

\end{appendix}

\end{document}